\documentclass[12pt]{iopart}
\usepackage{amssymb,amsfonts,dsfont,float,graphics,epsfig,epstopdf,color,verbatim,tabularx,bm,multirow}

\usepackage{iopams}  
\expandafter\let\csname equation*\endcsname\relax
\expandafter\let\csname endequation*\endcsname\relax
\usepackage{amsmath}
\usepackage{mathrsfs}
\usepackage{cite}
\usepackage{subfigure}
\usepackage{dcolumn}
\usepackage{dcolumn}% Align table columns on decimal point
\usepackage[mathlines]{lineno}
\usepackage[colorlinks,
            linkcolor=blue,
            urlcolor=blue,
            citecolor=blue]{hyperref}
\usepackage{changes}

\begin{document}

\title[]{Pinning-depinning transitions in two classes of discrete elastic-string models in (2+1)-dimensions}

\author{Yongxin Wu and Hui Xia*}
\address{School of Materials Science and Physics, China University of Mining and Technology, Xuzhou 221116, China}
\ead{hxia@cumt.edu.cn}

\vspace{10pt}
\begin{indented}
\item[]November 2023
\end{indented}

\begin{abstract}
The pinning-depinning phase transitions of interfaces for two classes of discrete elastic-string models are investigated numerically. In the (1+1)-dimensions, we revisit these two elastic-string models with slight modification to growth rule, and compare the estimated values with the previous numerical and experimental results. For the (2+1)-dimensional case, we perform extensive simulations on pinning-depinning transitions in these { discrete models with quenched disorder}. For full comparisons in the physically relevant spatial dimensions, we also perform numerically two distinct  universality classes, including the quenched Edwards-Wilkinson (QEW), and the quenched Kardar-Parisi-Zhang (QKPZ) equations with and without external driving forces. The critical exponents of these  {systems in the presence of quenched disorder} are numerically estimated. Our results show that the critical exponents satisfy scaling relations well, and these two discrete elastic-string models do not fall into the existing universality classes. In order to visually comparisons of these {discrete systems with quenched disorder} in the (2+1)-dimensional cases, we present surface morphologies with various external driving forces during the saturated time regimes. The relationships between surface morphologies, scaling exponents and  correlation length are also revealed.
\end{abstract}

%\maketitle

\section{\label{sec1}Introduction}

In recent decades, much attention has been attached to the pinning-depinning transition of a driven interface in quenched random media \cite{barabasi1995fractal,tang1992pinning,lee2000growth,reichhardt2022kibble,zapperi1998dynamics,derrida2014depinning,csenbil2019observation,park2012domain,brown2019skyrmion,Wiese2022,Doussal2003functional,Rosso2001Origin}, which has served for interpretation of many other physical phenomena such as immiscible displacements of fluids in porous media \cite{rubio1989self}, field-driven motion of domain walls in magnetic systems \cite{huse1985pinning}, vortices and flux lines in superconductors \cite{campi2015inhomogeneity,lee2000growth,reichhardt2022kibble,fisher1991thermal,chaturvedi2020critical,tauber1995coulomb,tauber1995interactions,dobramysl2014pinning}.

As a typical paradigm, the Kardar-Parisi-Zhang (KPZ) equation is  widely utilized in describing the evolution of surface and interface \cite{corwin2012kardar,sasamoto2010one,Priyanka2021role,Wiese2022}. Driven by thermal noise $\zeta (\vec{r},t)$ \cite{fisher1991thermal}, the KPZ equation can be written as \cite{barabasi1995fractal}
\begin{equation}\label{eq1}
%\begin{equation}
    \frac{\partial h(\vec{r},t)}{\partial t}=\nu \nabla ^2 h(\vec{r},t)+\frac{\lambda}{2}(\nabla h)^2+ \zeta (\vec{r},t),
\end{equation}
where  $h(\vec{r},t)$ is the interface height with position $\vec{r}$ at time $t$, $\nu$ represents the surface tension smoothing the interface, and the nonlinear term with $\lambda$ reflects lateral growth.

However, 
%\deleted{the quenched noise $\eta (\vec{r},h)$ usually wins the competition with thermal noise in disordered medium, and plays a more important role in the pinning-depinning transition,}
in a heterogeneous medium, Eq. (\ref{eq1}) can not describe effectively the interface evolution. To do this, the simplest way is to substitute thermal noise $\zeta (\vec{r},t)$ with quenched noise $\eta (\vec{r},h)$ \cite{Parisi1992on,ramasco2001interface,huergo2014dynamic} generated by the quenched disorder \cite{campi2015inhomogeneity,fisher1991thermal,dotsenko1995critical,mardanya2018dynamics,pan2020non,deutschlander2013two,Wiese2022}. 
Thus, Eq. (\ref{eq1}) becomes the quenched KPZ (QKPZ),
\begin{equation}\label{eq2}
     \frac{\partial h(\vec{r},t)}{\partial t}=f+\nu \nabla ^2 h+\frac{\lambda}{2}(\nabla h)^2+\eta (\vec{r},h).
\end{equation}
Here $f$ is the external driving force, and the quenched noise $\eta (\vec{r},h)$ is regarded as Gaussian noise with zero mean, satisfying \cite{barabasi1995fractal}
\begin{equation}\label{eq3}
    \left\langle \eta(\vec{r},h)\eta(\vec{r'},h') \right\rangle=\delta^d(\vec{r}-\vec{r'})\Delta(h-h'),
\end{equation}
where the brackets $\langle \dots \rangle$ denote the sample average. Many studies have been devoted to the scaling behavior of the QKPZ model in the (1+1)-dimensional case \cite{lee2005depinning,song2007discrete,jeong1999facet,huergo2014dynamic}. However,  the investigation of the (2+1)-dimensional case is not sufficient, despite its physical relevance. For instance, when a cubic sponge contacts a water surface, the phenomenon of fluid propagation within the sponge driven by capillary forces, and the phenomenon of {two-dimensional}  magnetic domain wall evolving in a disordered {three-dimensional} ferromagnet \cite{Alava1996Disorder} can be described by (2+1)-dimensional  models { in the presence of quenched disorder}  \cite{barabasi1995fractal}. Therefore, we  focus on the pinning-depinning transition of the (2+1)-dimensional   {discrete models with quenched disorder} in the current work.

First, we briefly introduce the evolution of elastic strings and the pinning-depinning transition \cite{duemmer2007depinning,biswas2011depinning,tanguy1998individual,Rosso2001Origin}. In {these discrete models we study} (details in SEC.\ref{sec2}), an external driving force $f$ is applied to an elastic-string evolving in a medium with quenched disorder. While the interface is driven by $f$, it is also hindered by the pinning force $f_p$ generated by quenched disorder. In addition, there is another force called elastic force $f_{el}$, which is originated by the surrounding elements owing to elastic stretching of the string. When an element falls behind the surrounding ones, the elastic force acts as a driving force. Conversely, when the element exceeds the surrounding ones, the elastic force acts as a resistance. In other words, the elastic force tends to smooth the interface, whose function is similar to that of the diffusion term $\nu \nabla ^2 h$ in the QKPZ model. If the driving force $f$ is too small, the interface may be pinned by the quenched disorder, i.e., the long-term velocity of interface is zero. Inversely, when the driving force increases until it exceeds a threshold $f_c$, the interface will be driven through the medium with a non-zero velocity, i.e., depinning. Near $f_c$ , many interesting properties deserve to be studied, and various scaling exponents characterize the pinning-depinning transitions. 

 {In this work,} we are committed to the improvements and generalizations on the  {elastic-string models with quenched disorder} \cite{biswas2011depinning} in the  case of (1+1)- and (2+1)-dimensions, and our numerical results obtained here are in agreement with some existing experiments \cite{Bouchaud1990Fractal,Maaloy1992Experimental,Elisabeth1997Scaling}. For full comparisons between our results and the existing  {theoretical predictions}, the QKPZ equation is also selected to investigate numerically as the well-known universality class.  Actually, when $\lambda \to 0$, QKPZ reduces to the quenched Edwards-Wilkinson (QEW) \cite{edwards1982surface,barabasi1995fractal}. We also compare our results on QEW with existing discrete models that belong to the QEW universality class \cite{song2008discrete}.

The paper is organised as follows: Firstly, we introduce models and numerical methods to quantitatively describe the {discrete elastic-string systems with quenched disorder}. Then, we present our numerical results from two classes of discrete elastic-string models in the (1+1)- and (2+1)-dimensions, respectively, and compare them with those from QEW and QKPZ for the (2+1)-dimensional cases. Finally, discussions and conclusions are provided.

\section{\label{sec2} Models and basic concepts}
Here we briefly introduce two typical elastic-string models originally proposed in \cite{biswas2011depinning}. We slightly modify them in (1+1)-dimensions, and then generalize  to (2+1)-dimensional case. As a comparison, the discretized scheme of Eq. (\ref{eq2}) in (2+1)-dimensions is also introduced. Then we present some basic concepts used in the subsequent numerical simulations.
\subsection{Model I}
Model I follows the evolving equation,
\begin{equation}\label{eq4}
    \frac{\partial h(\vec{r}, t)}{\partial t}=\left|f_{el}\right|\hat{S}(\vec{r},t)+\eta\left(\vec{r},h\left(\vec{r},t\right)\right)+f,
\end{equation}
where $h(\vec{r},t)$ is the height of element with coordinate $\vec{r}$ at time $t$, $f$ is the external driving force. The expression for $\hat{S}(\vec{r},t)$ can be interpreted as the description on elastic force in SEC. \ref{sec1}. Specifically, $\hat{S}(\vec{r},t)=sgn(h(x-1,t)-2h(x,t)+h(x+1,t))$ for the (1+1)-dimensions, and $\hat{S}(\vec{r},t)=sgn(h(x,y-1,t)+h(x,y+1,t)$ $+ h(x-1,y,t)+h(x+1,y,t)-4h(x,y,t))$ for the (2+1)-dimensional case. $\eta(\vec{r},h(\vec{r},t))$ represents the quenched disorder in medium, which is similar to the impurities distributed in paper towels in the (1+1)-dimensions or the disorder in {three-dimensional} ferromagnet\cite{Alava1996Disorder}. One can observe that the characteristic of quenched disorder is that the disorder at a specific position in the medium does not change with time. We set $\eta$ to be uniformly distributed on the interval [-0.5,0.5]. In comparison with the setting in \cite{biswas2011depinning}, where the noise has a $50\%$ probability of 0 and the other $50\%$ probability is uniformly distributed in a negative interval, we believe our setting is more general. In the end, we obtain different results from those presented in \cite{biswas2011depinning}.
%\deleted{, allowing $\eta$ to act as a resistance to the interface moving forward.}

The magnitude of the elastic force in the (1+1)-dimensional case is determined by the following formula,
\begin{equation}\label{eq5}
    |f_{el}|=\frac{1}{L}\sum_x\left\{\sqrt{\left[h(x,t)-h(x+1,t)\right]^2+1}-1\right\},
\end{equation}
and the corresponding (2+1)-dimensional form reads,
%\begin{multline}\label{eq6}
\begin{equation}\label{eq6}
\begin{aligned}
   & |f_{el}|=\frac{1}{L^2}\sum_{x,y}\biggl\{\sqrt{\left[h(x,y,t)-h(x+1,y,t)\right]^2+1}\\
   & +\sqrt{\left[h(x,y,t)-h(x,y+1,t)\right]^2+1}-2\biggr\}.
\end{aligned}
%\end{multline} 
\end{equation} 

In order to investigate the pinning-depinning transition of Model I, we will directly perform numerical simulations on the finite-difference (FD) form \cite{amar1990numerical,Z1993Dynamics,2006Depinning,ramasco2001interface} of Eq. (\ref{eq4}), which reads,
\begin{equation}\label{eq7}
    h(\vec{r},t+\Delta t)= h(\vec{r},t)+g_1(\vec{r},t)\Delta t,
\end{equation}
where $g_1(\vec{r},t)= |f_{el}|\hat{S}(\vec{r},t)+\eta\left(\vec{r},\widetilde{h}(\vec{r},t)\right)+f$, $\widetilde{h}$ is the integer part of $h(\vec{r},t)$, and $\Delta t$ is time step. Fluctuations in the value of $\Delta t$ within a reasonable range do not affect the results.
Interestingly, one could also perform the following discrete microscopic form, i.e. the cellular automaton model \cite{leschhorn1993Interface,leschhorn1996anisotropic} to investigate numerically the scaling properties of the elastic-string systems: $ h(\vec{r},t+\Delta t)=h(\vec{r},t)+1$ if $g_1(\vec{r},t) > 0$, $ h(\vec{r},t+\Delta t)=h(\vec{r},t)$ if $g_1(\vec{r},t) \le 0$.
%\begin{equation}\label{eq7a}
%    h(\vec{r},t+\Delta t)=
%    \begin{cases}
%    h(\vec{r},t)+1,&\text{if}\quad g_1(\vec{r},t) > 0,\\
%    h(\vec{r},t),&\text{if}\quad g_1(\vec{r},t) \le 0.
%    \end{cases}
%\end{equation}
Previous studies \cite{mukerjee2022depinning,Mukerjee2022DepinningII} show that the well-known class of cellular automaton model introduced by Tang and Leschhorn \cite{tang1992pinning,leschhorn1993Interface,leschhorn1996anisotropic} could map exactly to the QKPZ universality class, at least for (1+1)- and (2+1)-dimensions.

\subsection{Model II}
For Model II, we set the magnitude of the elastic force $F$ to be constant. Firstly, we revisit this {quenched} model, and could reproduced some numerical results from Ref. \cite{biswas2011depinning}. Building upon this, we employ quenched noise $\eta\left(\vec{r},h(\vec{r},t)\right)$ distributed uniformly in a symmetric interval. In contrast to \cite{biswas2011depinning}, the noise is not only a function of 
$\vec{r}$, but also a function of $h(\vec{r},t)$. This noise acts as a real quenched disorder generated by impurity in heterogeneous materials where phenomena like crack front propagation \cite{ponson2010crack,alava2006statistical} occur. The evolving equation for height can be written as:
\begin{equation}\label{eq8}
    \frac{\partial h(\vec{r}, t)}{\partial t}=F \hat{S}(\vec{r}, t)+\eta(\vec{r}, h(\vec{r}, t))+f,
\end{equation}
with constant $F$.

Similar to Model I, we need to discretize Eq. (\ref{eq8}), 
and the corresponding FD scheme reads,
\begin{equation}\label{eq9}
    h(\vec{r},t+\Delta t)=h(\vec{r},t)+g_2(\vec{r},t)\Delta t,
\end{equation}
where $g_2(\vec{r},t)=F\hat{S}(\vec{r},t)+\eta\left(\vec{r},h(\vec{r},t)\right)+f$. The meanings of $\hat{S}$ and $f$ remain consistent with the corresponding variables in Model I.
Notedly, when performing numerical simulations based on the corresponding cellular automaton version of Model II, the height $ h(\vec{r},t)$ is incremented by one unit if the value of $g_2(\vec{r},t)$ is greater than zero, that is, $ h(\vec{r},t+\Delta t)=h(\vec{r},t)+1$ when $g_2(\vec{r},t) > 0$; otherwise, no action takes place. Based on the rules, one could obtain the scaling properties near the critical threshold.

%Notedly, when performing the numerical simulations based on the corresponding cellular automaton version of Model II, that is, 
%\begin{equation}\label{eq9b}
%    h(\vec{r},t+\Delta t)=
%    \begin{cases}
%    h(\vec{r},t)+1,&\text{if}\quad g_2(\vec{r},t) > 0,\\
%    h(\vec{r},t),&\text{if}\quad g_2(\vec{r},t) \le 0.
%    \end{cases}
%\end{equation}
%One could obtain the similar scaling properties from Eq. (\ref{eq9b}) with that of the FD version near the critical threshold.}

\subsection{(2+1)-dimensional QKPZ}
We numerically integrate Eq. (\ref{eq2}) in (2+1)-dimensions using the FD numerical scheme \cite{amar1990numerical,Z1993Dynamics,2006Depinning,ramasco2001interface}. The height $h(x,y,t)$ is updated by the evolving discretized scheme,
\begin{equation} \label{eq10}
\begin{aligned}
   & h(x,y,t+\Delta t) =h(x,y,t)+\Delta t \biggl\{\nu \bigl[ h(x+1,y,t)\\ 
   & +h(x-1,y,t)+h(x,y+1,t)+h(x,y-1,t)-4h(x,y,t)\bigr]\\
   &+\frac{\lambda}{8} \bigl[h(x+1,y,t)-h(x-1,y,t)\bigr]^2\\
   & +\frac{\lambda}{8} \bigl[h(x,y+1,t)-h(x,y-1,t)\bigr]^2\\  
   & +\eta(x,y,\tilde{h}) +f\biggr\}.
\end{aligned}
\end{equation}

In the subsequent simulations, we set $\nu=1$, $\Delta t=0.01$. $\tilde{h}$ represents the integer part of height. Without loss of generality, $\eta(x,y,\tilde{h})$ can be set to be uniformly distributed in [-4, 4] \cite{song2007discrete}. The simulation based on time evolution of Eq. (\ref{eq10}) begin from $h(x,y,0)=0$ with periodic boundary condition.

\subsection{Basic concepts}

\begin{figure}[!htbp]
 \centering
         \begin{minipage}[t]{0.5\textwidth}
            \centering          %子图居中
            \includegraphics[width=90mm]{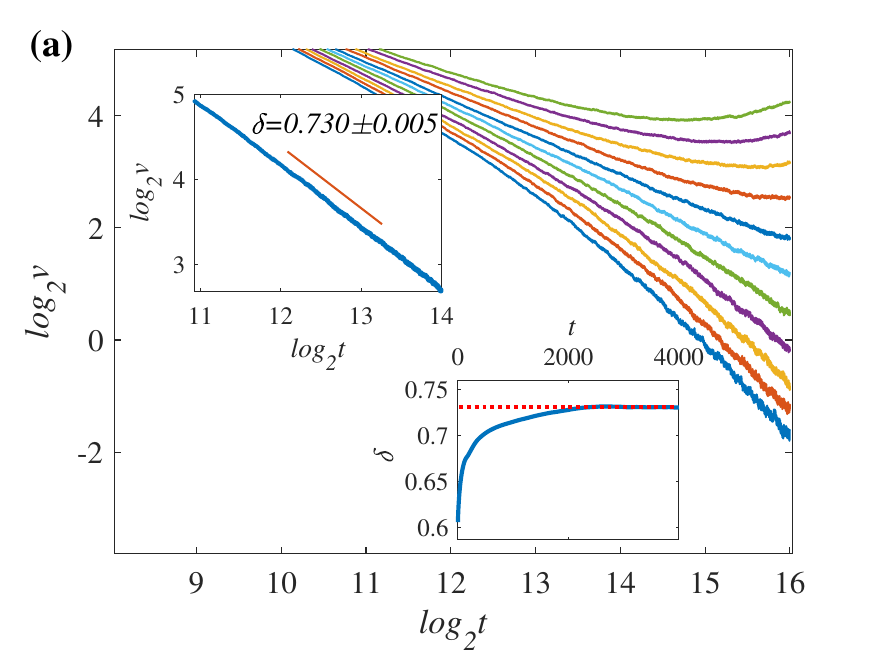}   
        \end{minipage}
   \\
        \begin{minipage}[t]{0.5\textwidth}
            \centering          %子图居中
            \includegraphics[width=90mm]{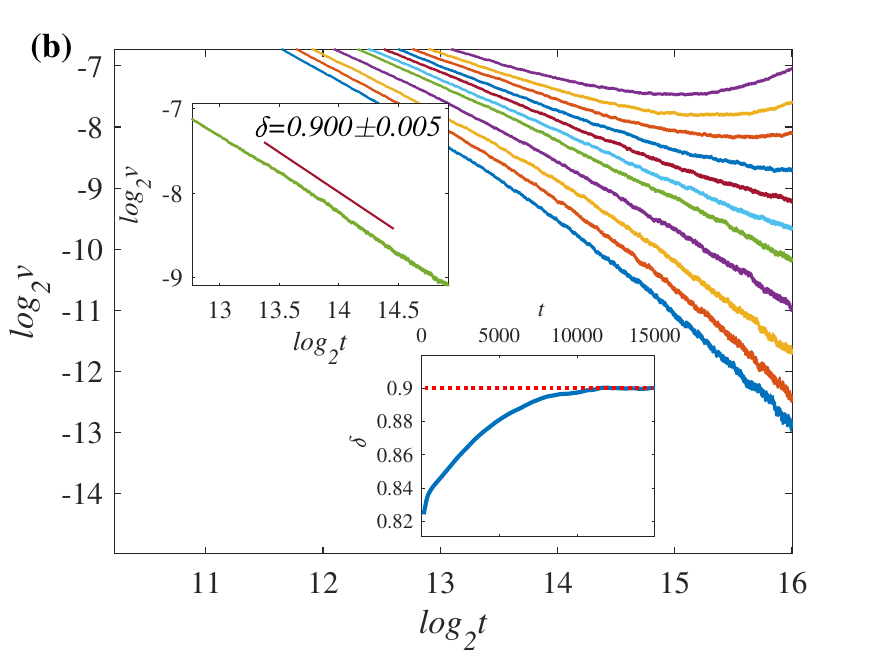}  
        \end{minipage}
\caption{Log-log plots of average velocity $v$ as a function of time $t$ in Model I with $\Delta t=0.01$. (a) The driving force $f=0.180,0.176,\dots,0.140$ from the top to the bottom with the system size $L=1024$ in the (1+1)-dimensions. (b) $f=0.133,0.131,\dots,0.121,0.117,0.113,\dots, 0.105$ from the top to the bottom with system size $L\times L=110\times110$ in the (2+1)-dimensions. Top inset shows $v$ versus $t$ near the critical thresholds: (a) $f_c \approx0.160$ and (b) $f_c \approx0.121$. Bottom inset exhibits the estimated value of $\delta$ as a function of time $t$.}
\label{fig1}
\end{figure}

In order to quantitatively depict the pinning-depinning transition, certain characteristic physical quantities need to be introduced. During the time evolution, the average velocity of moving interface can be defined as
\begin{equation}\label{eq11}
    v(t)=\frac{1}{L}\sum_{i=1}^{L}{\left(h(i,t)-h(i,t-1\right))}.
\end{equation}

At the critical point, the average velocity of interface follows,
\begin{equation}\label{eq12}
    v(t) \sim t^{-\delta},
\end{equation}
where $\delta$ is  a critical exponent \cite{Hinrichsen2000non,lee2005depinning}. This is the important criterion for determining the critical state, and similar scaling forms related to the velocity of interface can also be found in the following references \cite{Ferrero2013Nonsteady,dickman2000interface,Ferrero2013Numerical,Albano2011Study,Foini2013Static}. Once the critical point is successfully identified, several scaling exponents related to the pinning-depinning transition can be calculated.

The interface width, denoted as the mean square fluctuation, assists in characterizing the roughness properties of the interface,
\begin{equation}\label{eq13}
    W(L,t)=\frac{1}{\sqrt{L}}\langle \sum_{i}\left[h(i,t)-\langle h(t)\rangle \right]^2\rangle^{1/2},
\end{equation}
which follows the scaling ansatz \cite{barabasi1995fractal},
\begin{equation}\label{eq14}
    W(L,t) \sim
    \begin{cases}
   t^\beta \quad &\text{if} \quad t \ll t_\times\\
   L^\alpha\quad &\text{if} \quad t \gg t_\times,
    \end{cases}
\end{equation}
where $t_\times \sim L^z$. $\beta$ is the growth exponent, $z$ denotes the dynamic exponent and $\alpha$ is the roughness exponent.

One can expect $v\left(t\right) \sim W\left(L,t\right)/t \sim t^{\beta-1}$ in the critical state, implying $\beta +\delta=1$ \cite{lee2005depinning,kolton2006short,dickman2000interface,duemmer2007depinning,Ferrero2013Nonsteady}. To calculate the roughness exponent and dynamic exponent more precisely, we consider the height-height correlation function between two points with a distance $r$ \cite{siegert1996determining,jeong2002restricted,lopez1997super},
\begin{equation}\label{eq15}
    C(r,t)=\left\langle(h(x+r,t)-h(x,t))^{2}\right\rangle.
\end{equation}
The local roughness exponent $\alpha_{loc}$ can be obtained from the relation $C(r, t) \sim r^{2\alpha_{loc}} \quad(r \ll L)$. Normally, the relation $\alpha_{loc}=\alpha$ can be satisfied \cite{Ramasco2000Generic}. Further,  $C$, $\alpha$ and $z$ satisfy
\begin{equation}\label{eq16}
    C(r, t) \sim r^{2 \alpha} \mathscr{G}\left(r / t^{1 / z}\right),
\end{equation}
which can be used to obtain $z$ by data collapse. The scaling relation $z=\alpha/\beta$ should be satisfied \cite{barabasi1995fractal}.

With time evolution, the stationary velocity above the critical point follows \cite{barabasi1995fractal,lee2005depinning,le2020universal},
\begin{equation}\label{eq17}
    v\left(t\rightarrow\infty,f\right) \sim \left(f-f_c\right)^\theta,
\end{equation}
where $\theta$ is the velocity exponent. Interestingly, when $f=0$, one can achieve the critical point by adjusting the amplitude of the nonlinear coefficient. And there also exists a scaling ansatz above the critical $\lambda_c$ \cite{ramasco2001interface},
\begin{equation}\label{eq18}
    v\left(t\rightarrow\infty,\lambda\right) \sim \left(\lambda-\lambda_c\right)^{\theta'}.
\end{equation}
In order to further obtain the critical exponent near the critical point, we consider the  universal scaling ansatz as follows \cite{Hinrichsen2000non,kim2006depinning,biswas2011depinning},
\begin{equation}\label{eq19}
    v\left(f,t\right) \sim t^{-\delta}\mathscr{F}\left(t\left|f-f_c\right|^\nu\right),
\end{equation}
where $\nu$\footnote[1]{{In Refs. \cite{Hinrichsen2000non,song2008discrete}, $\nu$ is written as $\nu_{\parallel}$ or $\nu_t$, which is the critical exponent for the correlation in the time direction. Another critical exponent is $\nu_{\perp}$ or $\nu_x$, which is related to the lateral correlation. And the dynamical exponent $z=\nu_t/\nu_x$. Despite some differences in the symbols and expressions of these scaling exponents, the scaling relation  $\delta=\theta/\nu$  here is fundamentally compatible with those in Refs. \cite{kolton2006short,Ferrero2013Nonsteady}, for example.}} is the exponent related to the diverging correlation time, and correspondingly the scaling relation $\delta=\theta/\nu$ \cite{Hinrichsen2000non,song2008discrete} is always satisfied. By analogy, when $f=0$, the scaling ansatz is replaced by
\begin{equation}\label{eq20}
    v\left(\lambda,t\right) \sim t^{-\delta'}\Lambda\left(t\left|\lambda-\lambda_c\right|^{\nu'}\right).
\end{equation}

%避免叙述与前人重复

\begin{figure}[!htbp]
 \centering
        \begin{minipage}[t]{0.5\textwidth}
            \centering          %子图居中
            \includegraphics[width=90mm]{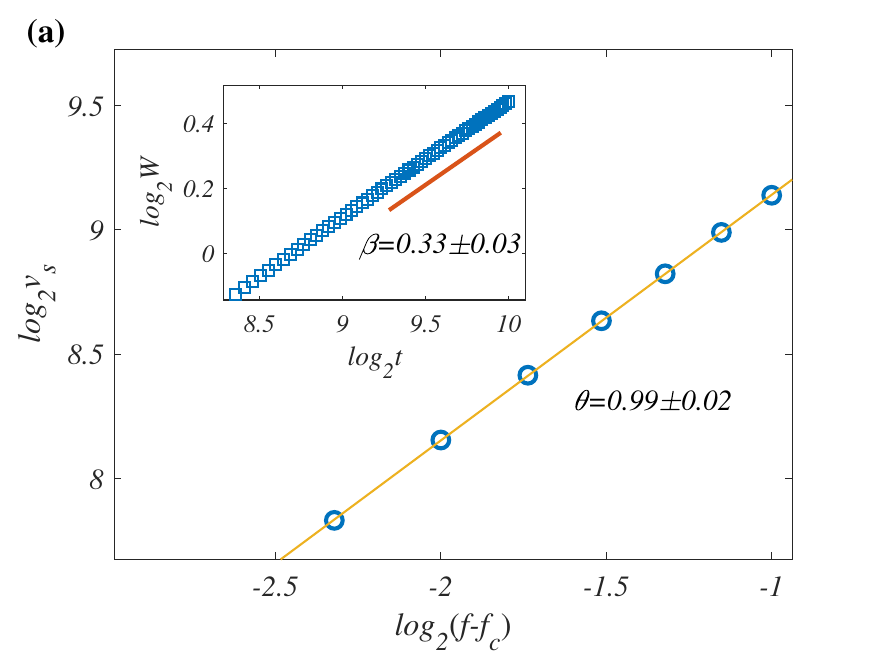}   
        \end{minipage}
  \\
        \begin{minipage}[t]{0.5\textwidth}
            \centering          %子图居中
            \includegraphics[width=90mm]{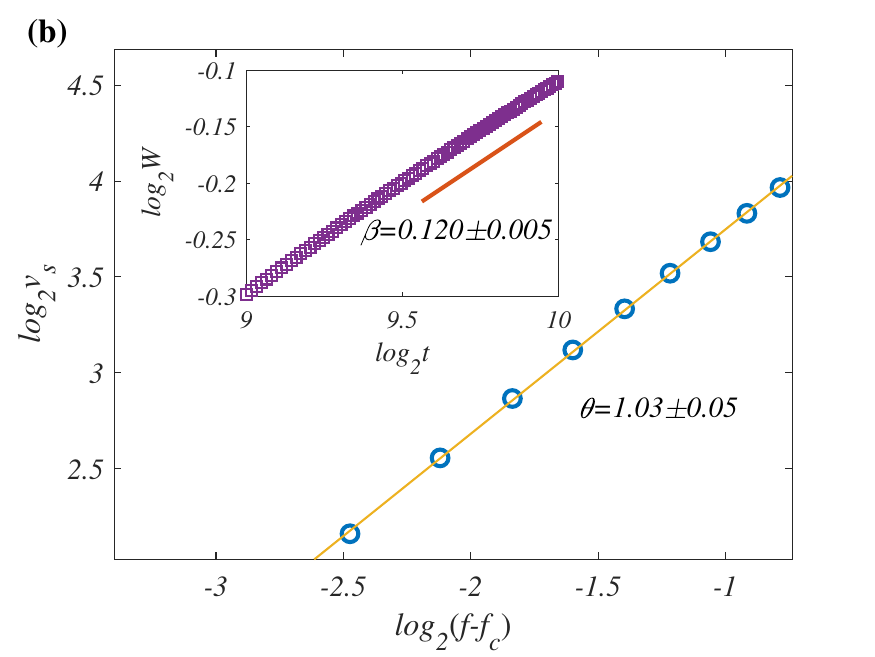}  
        \end{minipage}
\caption{Log-log plots of stationary velocity $v_s$ as a function of $f-f_c$ near the critical thresholds in Model I: (a) (1+1)-dimensional, and (b) (2+1)-dimensional cases. Insets show log-log plots of $W(L,t)$ as a function of $t$ at $f_c$, and the growth exponents are estimated correspondingly.}
\label{fig2}
\end{figure}

\section{\label{sec3}Numerical simulations and results}

\begin{figure*}[htbp]
 \centering
        \begin{minipage}[t]{0.48\textwidth}
            \centering          %子图居中
            \includegraphics[width=74mm]{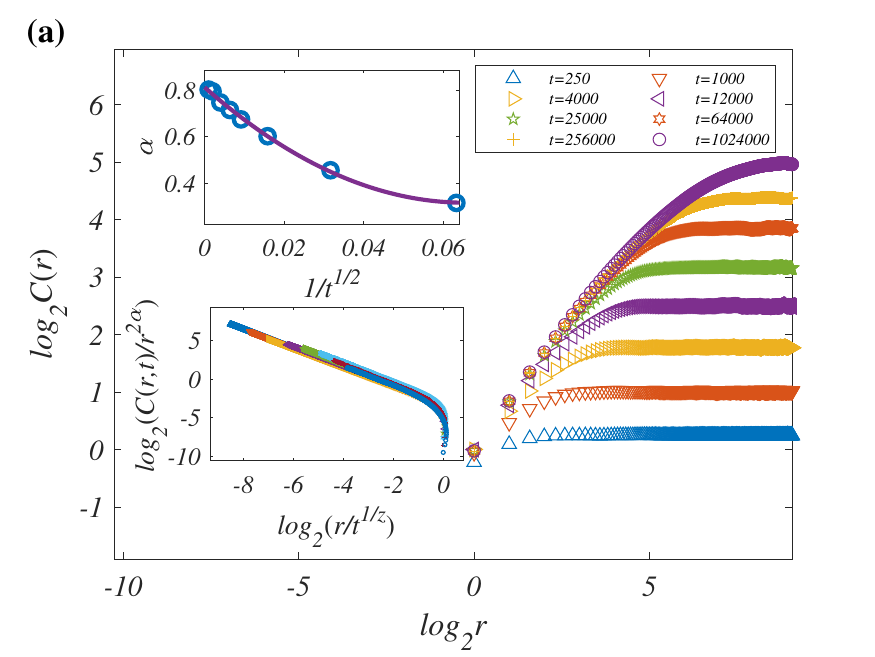} 
        \end{minipage}
         \begin{minipage}[t]{0.48\textwidth}
            \centering          %子图居中
            \includegraphics[width=74mm]{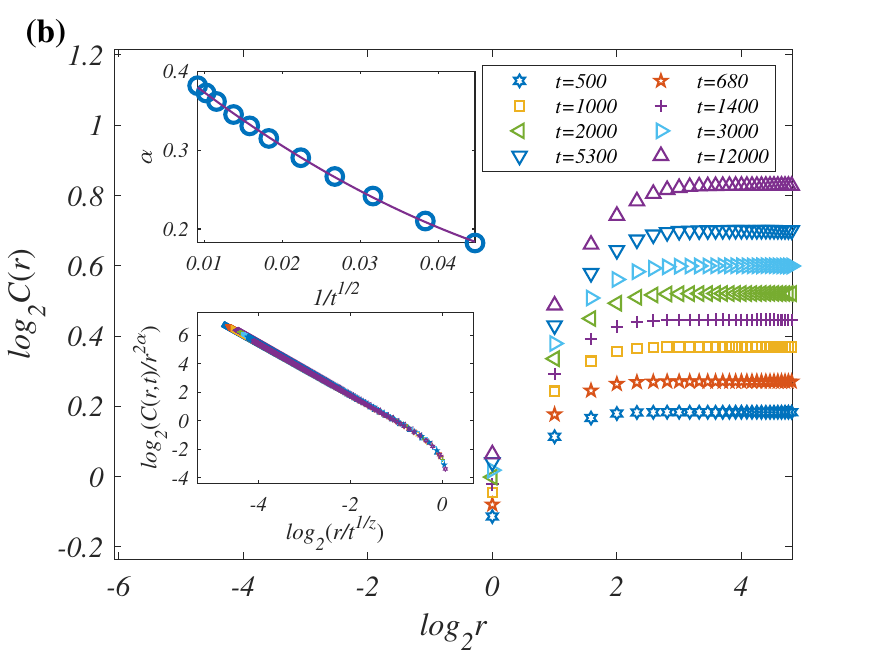} 
        \end{minipage}
  \\
         \begin{minipage}[t]{0.48\textwidth}
            \centering          %子图居中
            \includegraphics[width=74mm]{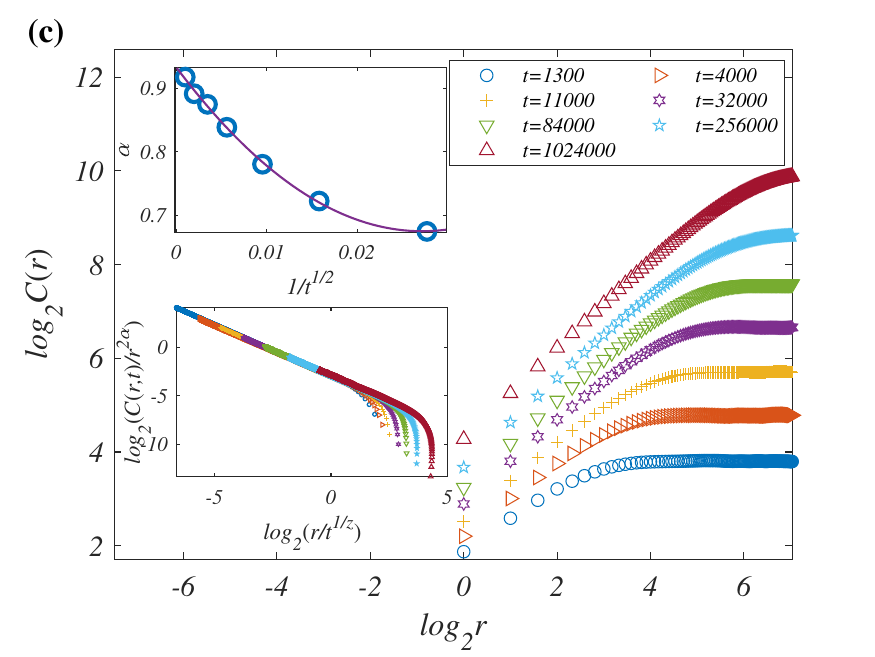} 
        \end{minipage}
        \begin{minipage}[t]{0.48\textwidth}
            \centering          %子图居中
            \includegraphics[width=74mm]{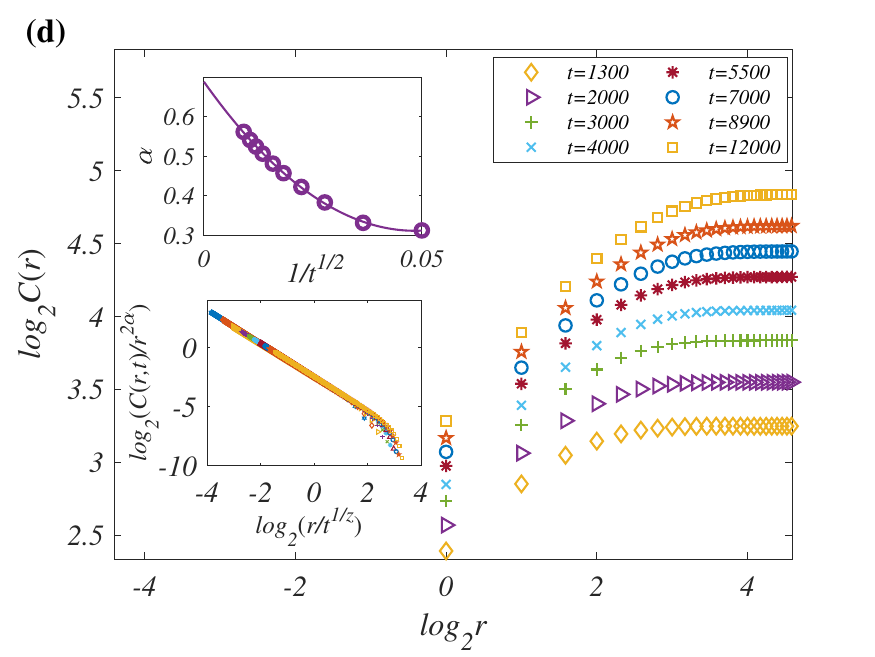} 
        \end{minipage}
     \caption{Log-log plots of the height-height correlation function $C(r, t)$ as a function of $r$ for { these two elastic-string} systems at different time: (a) the (1+1)- and (b) (2+1)-dimensional Model I with $\Delta t=0.01$, (c) the (1+1)- and (d) (2+1)-dimensional Model II with $\Delta t=0.05$. Here $L=4096$ for the (1+1)-dimensions and $L\times L=1024\times 1024$ for the (2+1)-dimensions are used. Top insets show $\alpha$ as a function of $1/t^{1/2}$ with quadratic polynomial fitting. Bottom insets exhibit data collapse with the effective estimated values of $\alpha$ and $z$.
     }
     \label{fig3}
\end{figure*}

\begin{figure}[!htbp]
 \centering
        \begin{minipage}[t]{0.5\textwidth}
            \centering          %子图居中
            \includegraphics[width=90mm]{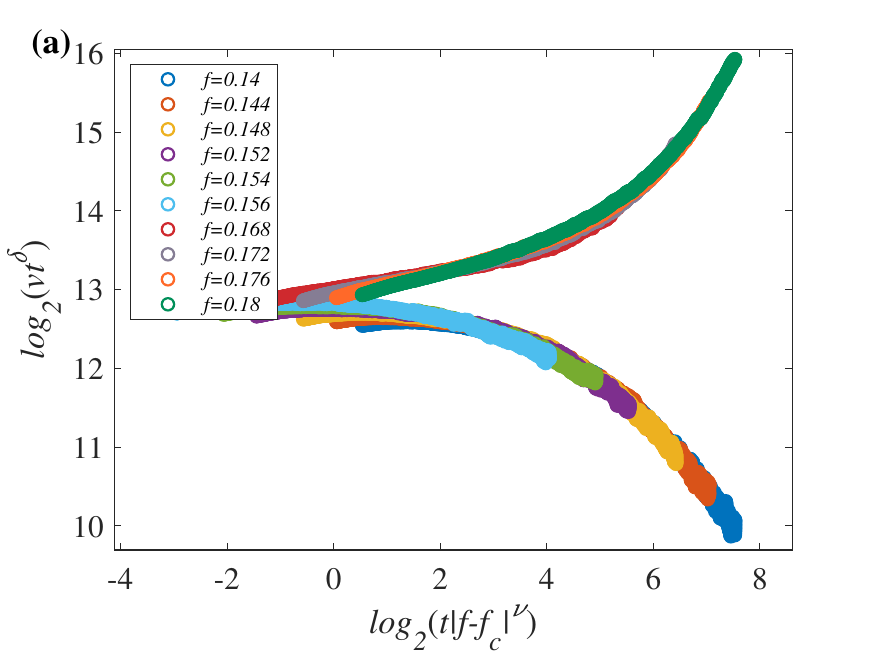} 
        \end{minipage}
   \\
        \begin{minipage}[t]{0.5\textwidth}
            \centering          %子图居中
            \includegraphics[width=90mm]{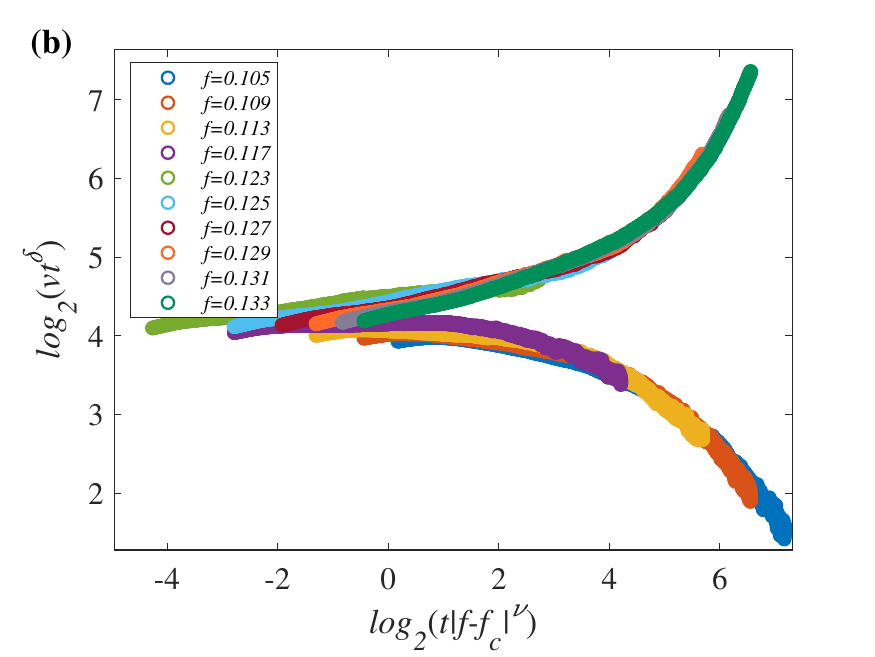}  
        \end{minipage}
    \caption{Near the critical thresholds, log-log plots of $vt^\delta$ as a function of $t|f-f_c|^\nu$ in Model I. The critical exponents are chosen in order to well fit data collapse: (a) $\delta=0.730$ and $\nu=1.50$ in the (1+1)-dimensions, (b) $\delta=0.900$ and $\nu=1.48$ in the (2+1)-dimensions. These lines for $f>f_c$ curve upward, and curve downward for $f<f_c$.}
    \label{fig4}
\end{figure}

\begin{figure}[!htbp]
 \centering
        \begin{minipage}[t]{0.5\textwidth}
            \centering          %子图居中
            \includegraphics[width=90mm]{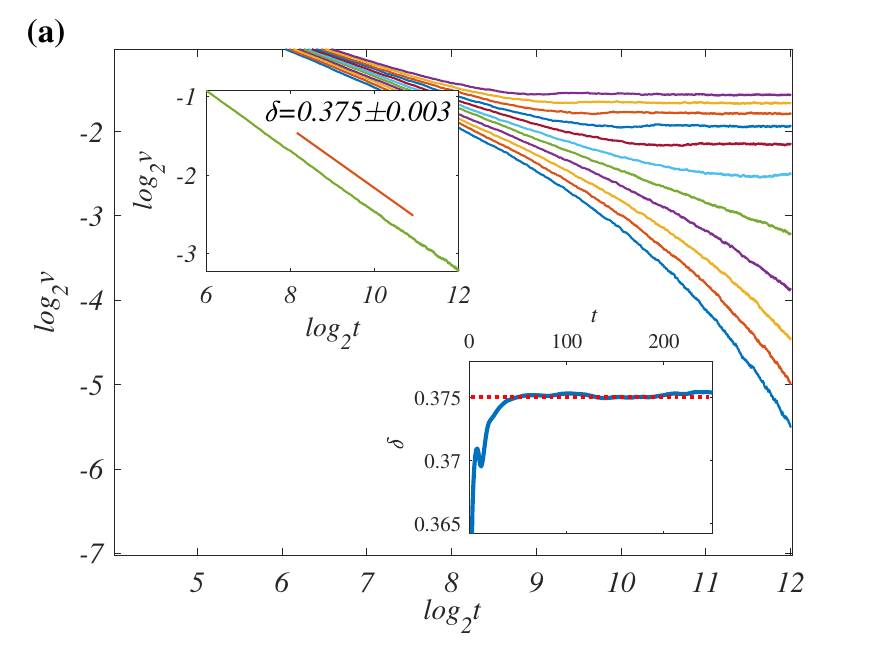} 
        \end{minipage}
  \\
        \begin{minipage}[t]{0.5\textwidth}
            \centering          %子图居中
            \includegraphics[width=90mm]{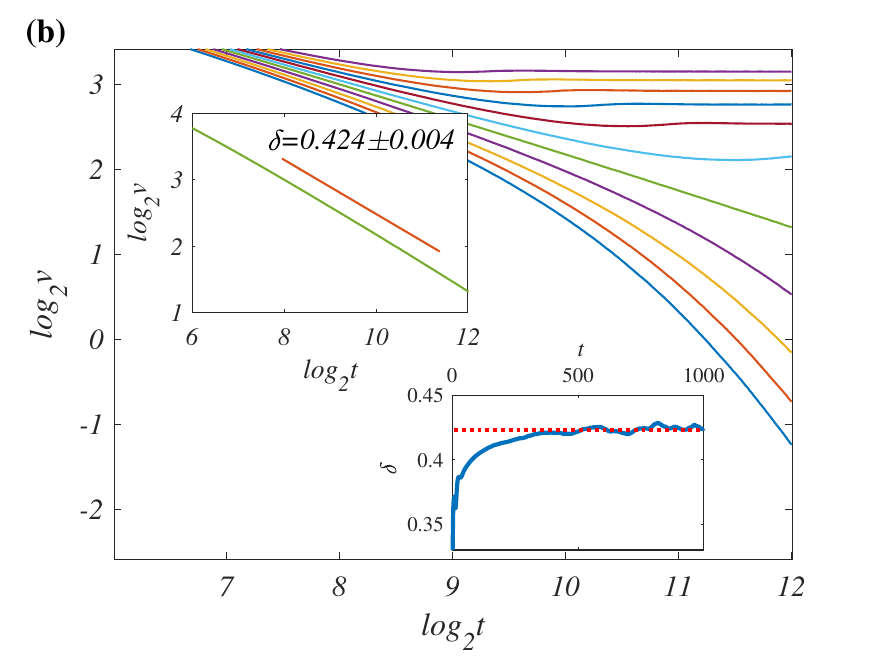}  
        \end{minipage}
     \caption{Log-log plots of $v$ as a function of time $t$ in Model II with $\Delta t=0.05$. The elastic force is fixed $F=0.6$, and the driving force $f=2.980,2.985,\dots,3.030$ are chosen from the bottom to the top. (a) shows the (1+1)-dimensional results with system size $L=1024$, and (b) exhibits the (2+1)-dimensional results with system size $L\times L=512\times512$. As is indicated in the insets, near $f_{c}$, the estimated values of critical exponent $\delta$ are obtained correspondingly.}
     \label{fig5}
\end{figure}

\begin{figure}[htbp]
    \centering
    \includegraphics[width=90mm]{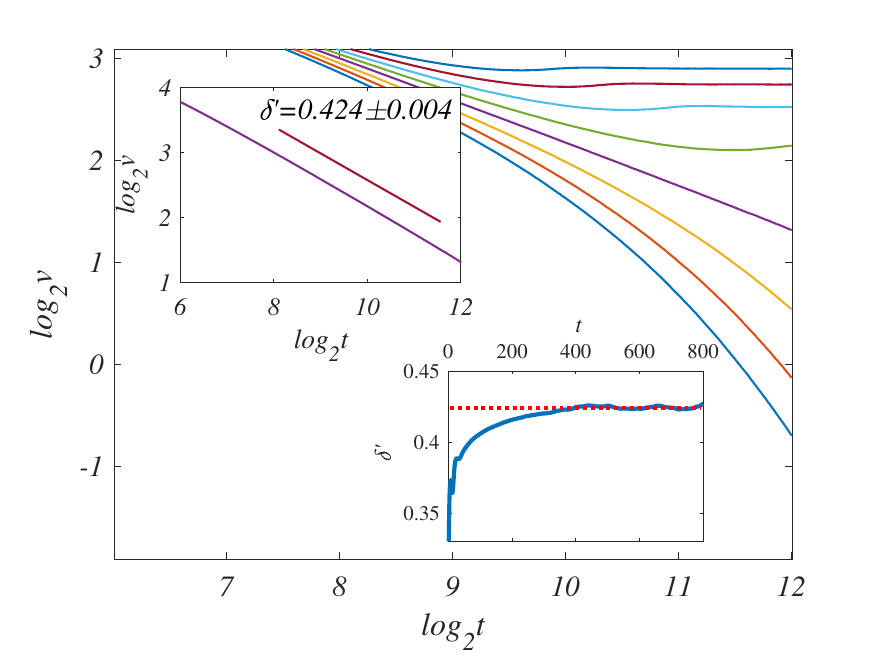}
    \caption{Log-log plot of $v$ as a function of time $t$ in the (2+1)-dimensional Model II with the external driving force $f=3.000$, $\Delta t=0.05$ and the elastic force $F=0.585,0.590,\dots,0.620$ from bottom to top. Top inset shows $v$ versus $t$ near the critical threshold $F_c \approx0.600$, and bottom inset exhibits the estimated value of $\delta'$ as a function of time $t$.}
    \label{fig6}
\end{figure}

\begin{figure}[htbp]
 \centering
        \begin{minipage}[t]{0.5\textwidth}
            \centering          %子图居中
            \includegraphics[width=90mm]{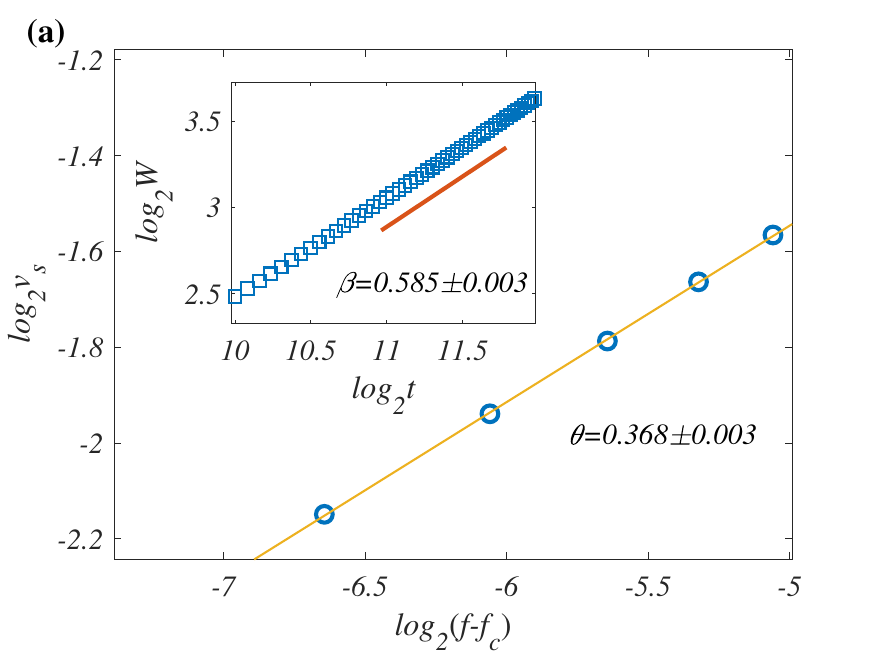} 
        \end{minipage}
 \\
        \begin{minipage}[t]{0.5\textwidth}
            \centering          %子图居中
            \includegraphics[width=90mm]{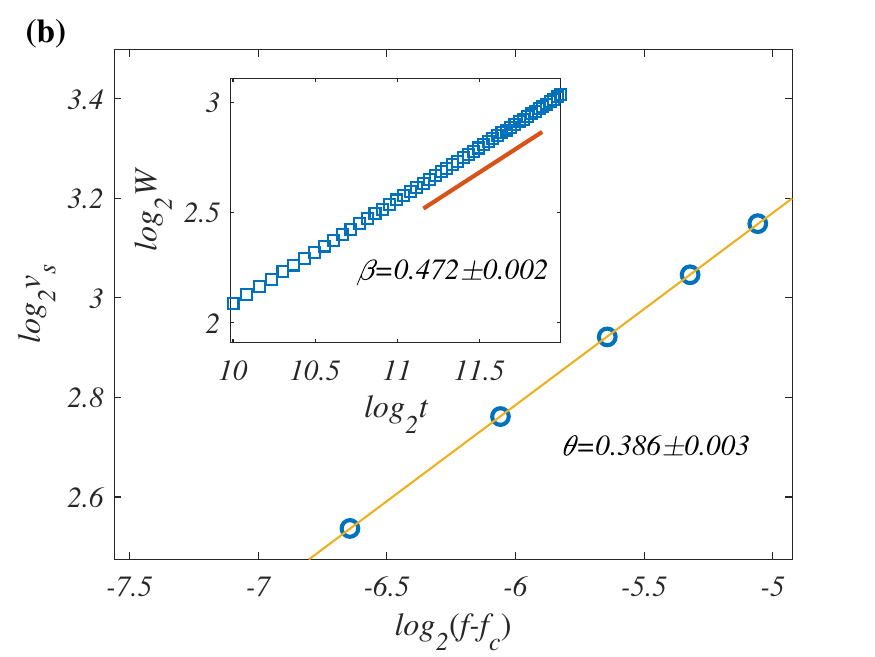}  
        \end{minipage}
     \caption{Log-log plots of stationary velocity $v_s$ as a function of $f-f_c$ in Model II. (a) $f_c\approx3.000, \theta=0.368\pm0.003$ in the (1+1)-dimensions. (b) $f_c\approx3.000, \theta=0.386\pm0.003$ in the (2+1)-dimensions. Insets show log-log plots of $W$ as a function of $t$ near $f_c$, and the values of growth exponent $\beta$ are estimated correspondingly.}
     \label{fig7}
\end{figure}

\begin{figure}[htbp]
 \centering
        \begin{minipage}[t]{0.5\textwidth}
            \centering          %子图居中
            \includegraphics[width=90mm]{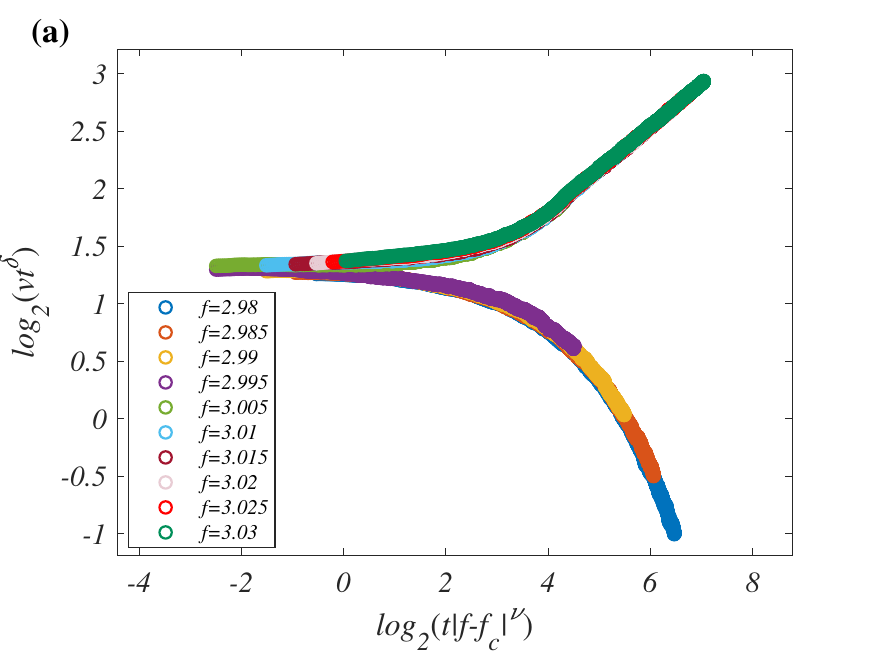} 
        \end{minipage}
  \\
        \begin{minipage}[t]{0.5\textwidth}
            \centering          %子图居中
            \includegraphics[width=90mm]{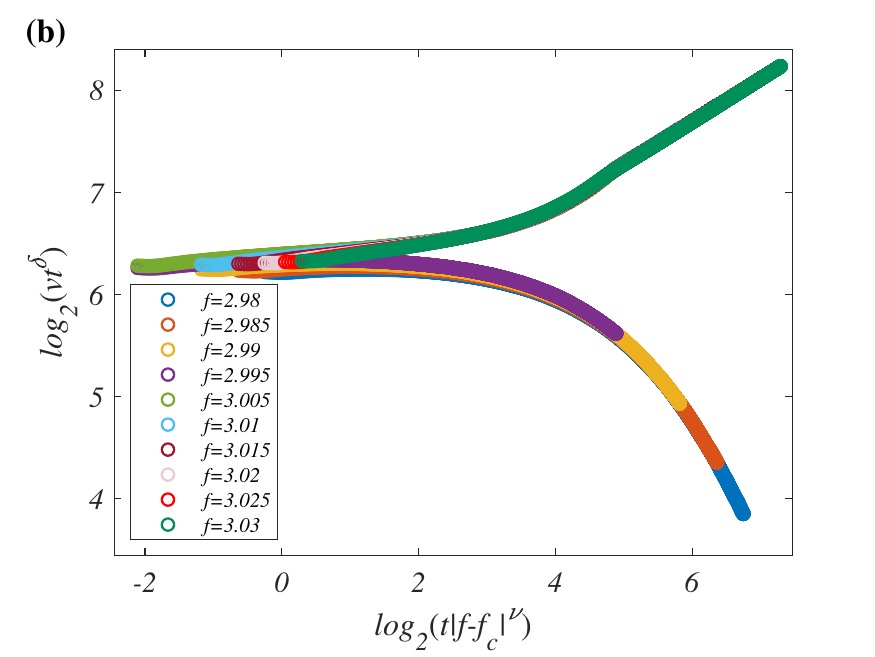}  
        \end{minipage}
     \caption{Log-log plots of $vt^\delta$ as a function of $t|f-f_c|^\nu$ in Model II. The critical exponents are chosen in order to well fit data collapse: (a) $\delta=0.375$ and $\nu=0.98$ in the (1+1)-dimensions, (b) $\delta=0.424$ and $\nu=0.93$ in the (2+1)-dimensions.}
     \label{fig8}
\end{figure}

\subsection{Model I}

In this subsection, we focus on numerical simulations based on the FD scheme, and estimate the values of the critical exponents at pinning-depinning transition in Model I. For this purpose, our priority is to identify the critical threshold. Considering that the magnitude of the elastic force is variable in this model, and $|f_{el}|$ follows Eq. (\ref{eq5}) or Eq. (\ref{eq6}), we need to tune the magnitude of driving force $f$ to induce the critical state. 
It should be noted in advance that the finite-size effects of {these elastic-string} models we simulated are relatively weak. However, the finite-time effects are not ignored. Thus we choose the appropriate system sizes to present our results in consideration of computational costs, and finite-time correlations are considered in dealing with some characteristic quantities accordingly.
In our numerical simulations, different system sizes are used: $L=1024$ in the (1+1)-dimensions, and $L\times L=110\times 110$ in the (2+1)-dimensional case. The power-law scaling of the average velocity $v$ versus time $t$  in both cases is shown in Fig. \ref{fig1}. For the pinned region ($f<f_{c}$), as the absolute value of curve slope increases, $v$ decays to 0 quickly. On the contrary, in the depinned region ($f>f_c$), $v$ decays following a power-law at early growth time, and then the curves bend upward. Only near $f_c$, $v$ shows a power-law scaling as $v \sim t^{-\delta}$. We obtain the critical exponents $\delta=0.730\pm0.005$ with $f_c\approx0.160$ in the (1+1)-dimensions, and $\delta=0.900\pm0.005$ with $f_c\approx0.121$ in the (2+1)-dimensions, as shown in the insets of Fig. \ref{fig1}.

When the critical threshold is determined, the scaling properties near the critical point can be investigated.  In the depinned region, we plot the saturated velocity $v_s$ as a function of $f-f_c$ in both (1+1)- and (2+1)-dimensional cases, as shown in  Figs. \ref{fig2} (a) and (b), respectively. Thus, we obtain the critical exponents $\theta=0.99\pm0.02$ in the (1+1)-dimensions, and $\theta=1.03\pm0.05$ for the (2+1)-dimensional case. As mentioned in SEC. \ref{sec2}, we exhibit the power-law scaling of surface roughness $W(L,t)$ as a function of time $t$. Typical curves of $W(L,t)$ at $f_c$ are shown in the insets of Fig. \ref{fig2}. The growth exponents $\beta=0.33\pm0.03$ and $\beta=0.120\pm0.005$ are obtained from the (1+1)- and (2+1)-dimensional cases, respectively. Our results show that $\delta+\beta\approx1.06$ in the (1+1)-dimensions, and $\delta+\beta\approx1.02$ in the (2+1)-dimensions, which implies that these critical exponents satisfy the scaling relation $\beta+\delta=1$ very well in these two dimensions. To further quantitatively describe the scaling properties of these {elastic-string models} at the critical threshold, we use the scaling relation of height-height correlation function $C(r,t)$ to determine other scaling exponents. Typical curves of $C(r,t)$ for these  elastic-string systems at different times are shown in Figs. \ref{fig3}. We observe that the scaling behavior of the height-height correlation function exhibits strong dependence on growth time, hence finite-time effects are not ignored any more. To suppress possible finite-time effects, we perform finite-time corrections to obtain the asymptotic value of $\alpha_{loc}$ when the growth time tends to infinity using a polynomial fitting form $\alpha_{loc}(t)\approx \alpha_{loc}+at^{-1/b}$.
The top insets show $\alpha$ as a function of $1/t^{1/b}$ with $a=1,b=2$. With quadratic polynomial fitting, the vertical intercepts of these insets are the corresponding estimated values of $\alpha_{loc}$. And then, based on scaling ansatz Eq. (\ref{eq16}), we also perform data collapses of $C(r, t)/ r^{2 \alpha}$ versus $ r / t^{1 / z}$ in the bottom insets, and the effective estimated values of $\alpha$ and $z$ are obtained correspondingly. 
 Detailed results are presented in Table \ref{tab}. And the scaling relation $z=\alpha/\beta$ is satisfied numerically in both (1+1)- and (2+1)-dimensions.
Furthermore, based on scaling ansatz (\ref{eq19}), we perform data collapses to determine the critical exponent $\nu$. As shown in Fig. \ref{fig4}, these data collapse into two lines nicely with $\delta=0.730$, $\nu=1.50\pm0.05$ in the (1+1)-dimensions, and $\delta=0.900$, $\nu=1.48\pm0.03$ in the (2+1)-dimensions. Thus, we verify that another scaling relation $\nu=\theta/\delta$ can be satisfied well in Model I.

\subsection{Model II}
Following the similar research steps to Model I, we first determine the criticality of Model II. Interestingly, the critical threshold here is not solely dependent on the external driving force $f$, but also depends on the magnitude of the elastic force $F$ between the elastic strings, so how to deal with the interplay between them becomes very important. 

Actually, through extensive numerical simulations, we find that $f$ and $F$ play driving roles in promoting the whole evolving process, while the quenched noise plays a resistance role (pinning force $f_p$). When the maximum resistance can be balanced with driving forces, the system reaches a critical state. Since the quenched noise is generated in a set of random numbers, the negative random number with the largest absolute value is the maximum resistance $f_{pmax}$. We can summarize the formula to determine the critical threshold through extensive numerical simulation:
\begin{equation}\label{eq21}
    f_{pmax}=f_c+F,
\end{equation}
where  $f_{pmax}$ is the amplitude of the random number interval $[-f_{pmax}, f_{pmax}]$, where the quenched noise is generated uniformly. In fact, through our simulations, we find that the value of the right end of the random number interval does not affect the correctness of Eq. (\ref{eq21}) and the estimated values of scaling exponents. Without losing generality, we set the elastic force $F=0.6$ and $f_{pmax}=3.6$. Obviously, according to Eq. (\ref{eq21}), the critical threshold {is} $f_c=3.000$, which will be verified numerically below.

As shown in Fig. \ref{fig5}, the scaling law of velocity versus time is similar to that of Model I. Near $f_c\approx3.000$, $v$ always decays in power-law $v \sim t^{-\delta}$, and one can obtain a nice straight line in the insets of Fig. \ref{fig5} with $\delta=0.375\pm0.003$ for the (1+1)-dimensions and $\delta=0.424\pm0.004$ for the (2+1)-dimensions. It is evident that numerical results agrees well with Eq. (\ref{eq21}). Even slight deviations of the driving force from the critical force result in significant changes in the curve. It is noteworthy that, based on the cellular automaton version of Model II, the critical threshold is highly sensitive to external parameters compared to the FD method.

%Furthermore, we investigate the sensitivity of interface growth to the changes in elastic forces. Contrary to Fig. \ref{model2_delta}, we try to fix the driving force $f\approx3.0000$ and change the value of $F$ near the critical elastic force $F_c\approx0.6$ obtained from Eq. (\ref{m2c}). We also observe the changes in trends of different lines simulated to Fig. \ref{model2_delta}, as shown in Fig. \ref{model2_changeF}. Comparing the two figures, it is easy to find that the growth process is sensitive not only to the changes in driving force $f$, but also to the changes in elastic force $F$.

Furthermore, we investigate the sensitivity of interface to the changes in the elastic force $F$ for the (2+1)-dimensional Model II. We fix the driving force $f=3.000$ and adjust the values of $F$ to observe the pinning-depinning transition, then the critical threshold $F_c\approx0.600$ is obtained, further confirming Eq. (\ref{eq21}). Figure \ref{fig6} illustrates the changes in trends of different lines by adjusting $F$, akin to those by adjusting $f$ depicted in Fig. \ref{fig5}.  A comparison between these two figures reveals that the pinning-depinning transition is not only sensitive to changes in the driving force, but also to changes in the elastic force. And the estimated value $\delta'=0.424\pm0.004$ coincides with that of $\delta$ within the error range.

Near $f_c$, the interface width follows power-law as Eq. (\ref{eq14}) with $\beta=0.585\pm0.003$ in the (1+1)-dimensions, and $\beta=0.472\pm0.002$ in the (2+1)-dimensions, as shown in the insets of Fig. \ref{fig7}. Hence, our numerical results satisfy the relation $\beta+\delta=1$ very well. Figures \ref{fig3} (c) and (d) depict $C(r,t)$ in the (1+1)-dimensions and the (2+1)-dimensions for Model II. The corresponding exponents $\alpha$ and $z$ are  presented in Table \ref{tab}, and the relation $z=\alpha/\beta$ is satisfied very well. It has been noted that the evolving process is sensitive to adjusting the magnitude of the driving force, which helps us to  calculate the velocity exponent more accurately. In the depinned region, we plot the saturated velocity $v_s$ as a function of $f-f_c$, and obtain the critical exponents $\theta=0.368\pm0.003$ for the (1+1)-dimensions and $\theta=0.386\pm0.003$ for the (2+1)-dimensions, as shown in Fig. \ref{fig7}. Furthermore, based on power-law scaling (\ref{eq19}), we obtain another scaling exponent $\nu=0.98\pm0.03$ for the (1+1)-dimensions and $\nu=0.93\pm0.02$ for the (2+1)-dimensions, as shown in Fig. \ref{fig8}. The scaling relation $\nu =\theta/\delta$ is also numerically verified.

\subsection{(2+1)-dimensional QKPZ and QEW equations}

\begin{figure}[!htbp]
    \centering
    \includegraphics[width=90mm]{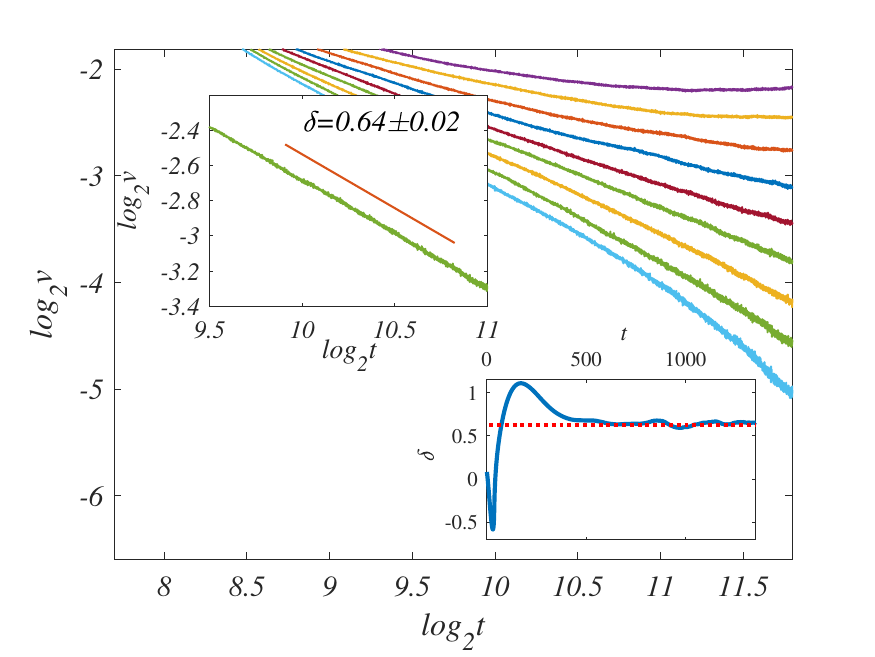}
    \caption{Log-log plot of $v$ as a function of time $t$ of the (2+1)-dimensional QKPZ with the nonlinear coefficient $\lambda=5$, $\Delta t=0.01$ and the external force $f=-0.66,-0.64,\dots,-0.50$ from bottom to top. The system size $L\times L=256\times256$. As is indicated in the insets, near $f_c\approx-0.60$, the estimated values of critical exponent $\delta$ is obtained correspondingly.}
    \label{fig9}
\end{figure}

\begin{figure}[!htbp]
    \centering
    \includegraphics[width=90mm]{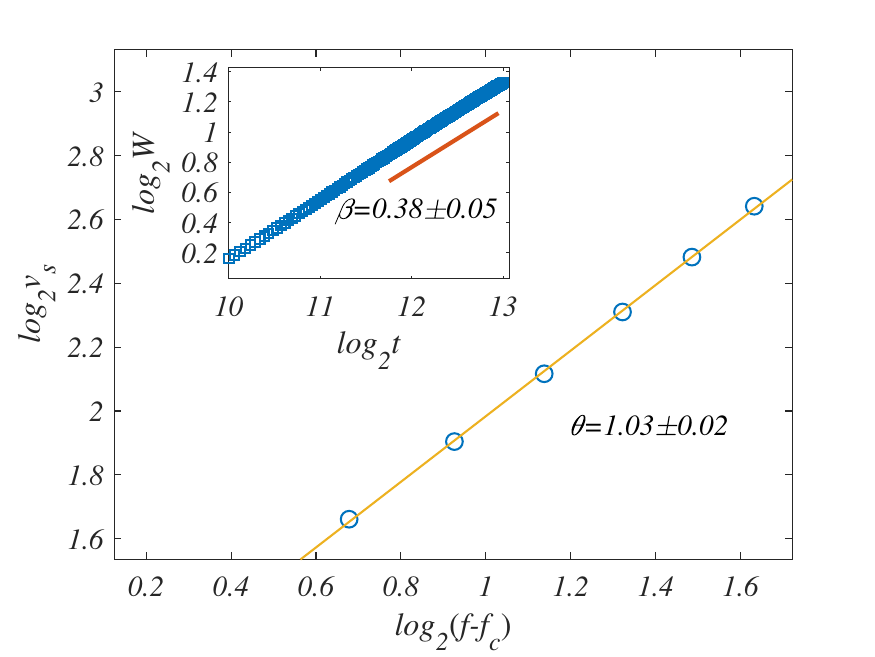}
    \caption{Log-log plot of stationary velocity $v_s$ as a function of  $f-f_c$ of the (2+1)-dimensional QKPZ with $f_c \approx-0.60$. Inset shows log-log plot of $W$ as a function of $t$ near $f_c$ and the estimated value of $\beta$. } 
    \label{fig10}
\end{figure}

\begin{figure}[htbp]
 \centering
       
         \begin{minipage}[t]{0.5\textwidth}
            \centering          %子图居中
            \includegraphics[width=90mm]{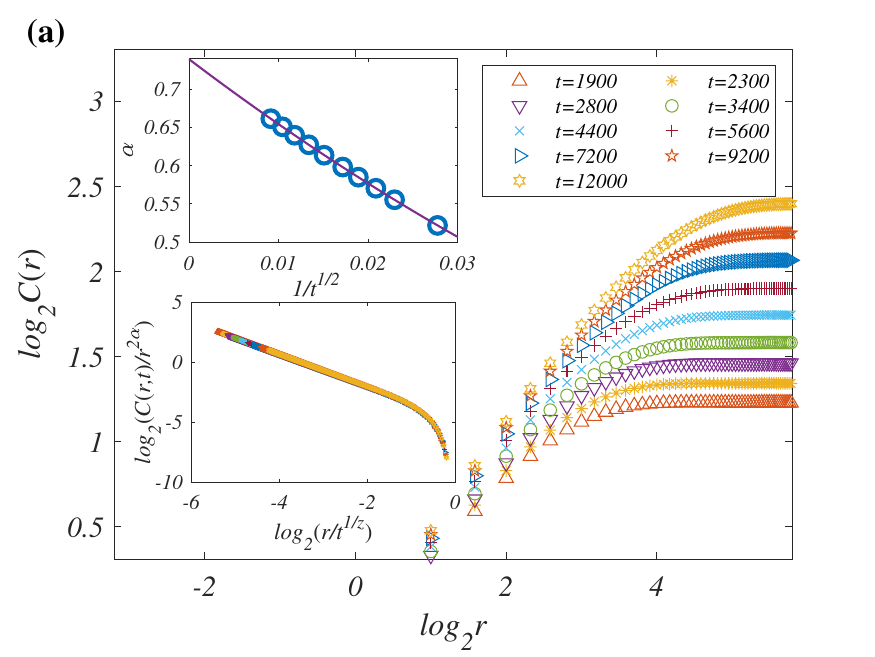} 
        \end{minipage}
    \\
        \begin{minipage}[t]{0.5\textwidth}
            \centering          %子图居中
            \includegraphics[width=90mm]{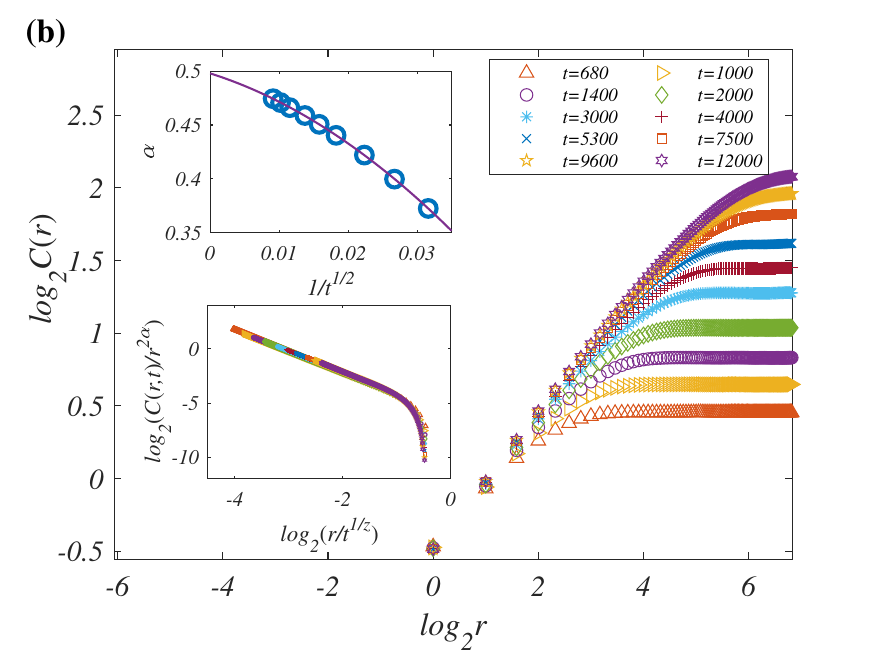} 
        \end{minipage}    
     \caption{Log-log plots of the height-height correlation function $C(r, t)$ as a function of $r$ for the quenched equations at different time: (a) the (2+1)-dimensional QEW and (b) the (2+1)-dimensional QKPZ. Here $L\times L=1024\times 1024$ and $\Delta t=0.01$ for the two systems are used. Top insets show $\alpha$ as a function of $1/t^{1/2}$ with quadratic polynomial fitting. Bottom insets exhibit data collapse with effective values of $\alpha$ and $z$.
     }
     \label{fig11}
\end{figure}

\begin{figure}[!htbp]
    \centering
    \includegraphics[width=90mm]{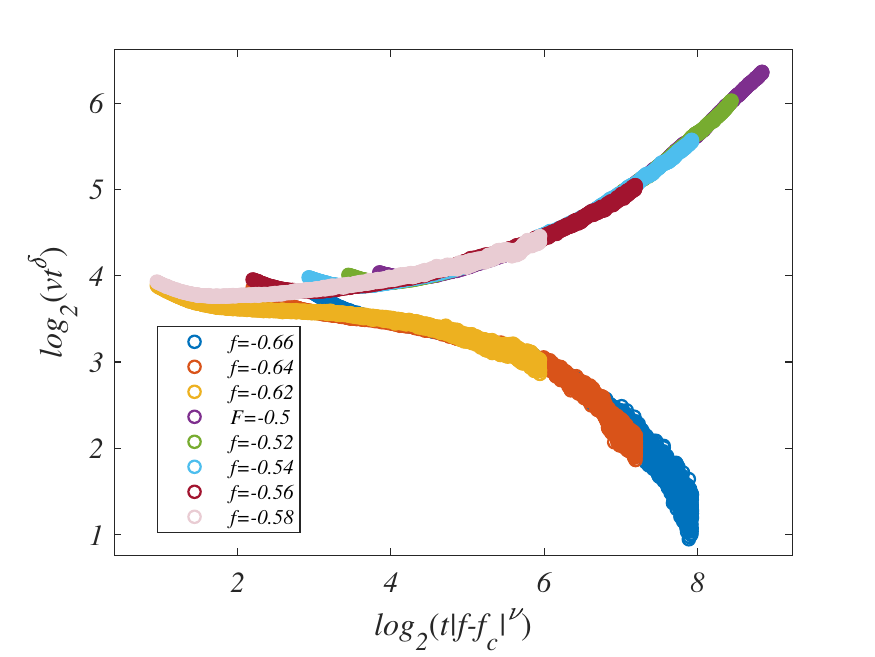}
    \caption{Log-log plot of $vt^{\delta}$ as a function of $t|f-f_c|^{\nu}$ of the (2+1)-dimensional QKPZ system. The critical exponents $\delta=0.64$ and $\nu=1.25$ are chosen in order to well fit data collapse.}
    \label{fig12}
\end{figure}

 In this subsection, we investigate numerically QKPZ and QEW equations as two well-known universality classes in the case of (2+1)-dimensions. The time evolution of interface is updated by Eq. (\ref{eq10}). We find that the coefficient of nonlinear term $\lambda$ is negatively correlated with the external parameter $f$. Actually, for any chosen $f$, the corresponding critical coefficient $\lambda_c$ can always be found to reach pinning-depinning transition. Firstly, we set $\lambda=5$ to ensure this system is far away from the EW universality class. Figure \ref{fig9} shows that the power-law scaling of average velocity $v$ versus time $t$ is similar to { those of the elastic-string }models mentioned above. In this case, we obtain the critical force  $f_c\approx-0.60$. Near $f_c$, $v$ always decays in power-law scaling $v\sim t^{-\delta}$ with $\delta=0.64\pm0.02$, as shown in the inset of Fig. \ref{fig9}. Meanwhile, we measure $W(L,t)$ as a function of $t$, and obtain $\beta=0.38\pm0.05$. Then, other scaling exponents near the critical point mentioned above are also calculated, as shown in Figs. \ref{fig10}-\ref{fig12}. The scaling relations $\beta+\delta=1$, $\nu=\theta/\delta$ and $z=\alpha/\beta$ are numerically verified. Recently, Mukerjee et al. \cite{mukerjee2022depinning,Mukerjee2022DepinningII} carried out extensive numerical and theoretical researches related to the QKPZ universality class, and obtained some scaling exponents. We find that our estimated values of the critical exponents including $z$ and $\alpha$ are in well agreement with those in these literatures.

%\deleted{Furthermore, we demonstrate explicitly the equivalence of the fixed nonlinear parameter with a small value and the fixed driving force simulations on the (2+1)-dimensional QKPZ model.  Two special cases are chosen: (i) QKPZ with a smaller $\lambda=2$ and $F_c\approx0.6$. (ii) QKPZ with $F=0$ and $\lambda_c\approx3.0$. Interestingly, we obtain the same scaling exponents within a certain error range from these two cases: $\delta=\delta'=0.520\pm0.005, \beta=0.425\pm0.005, \theta=\theta'=0.84\pm0.03$ and $\nu=\nu'=1.4\pm0.05$, implying that adjusting $F$ and $\lambda$ has equivalent effect on pinning-depinning transition, which has also been mentioned in \cite{ramasco2001interface} for (1+1)-dimensional cases. Notably, it is obvious that our results in these two special cases are deviated from those of QKPZ universality class in the (2+1)-dimensions. As a matter of fact, QKPZ in the absence of an external driving force or with small nonlinear coefficient, all could reduce gradually to EW universality class. }

Furthermore, we explicitly demonstrate the difference on two special cases of the (2+1)-dimensional QKPZ model: (i) QKPZ with the nonlinear coefficient $\lambda \to 0$. (ii) QKPZ with the external driving force $f \to 0$.
Firstly, considering that the nonlinear coefficient decreases gradually to $0$, the QKPZ system will crossover to QEW class \cite{edwards1982surface,barabasi1995fractal}. With 
a system size of $L\times L=256\times256$, we obtain that $ f_c\approx1.215, \delta=0.51\pm0.02, \beta=0.45\pm0.01, \alpha=0.74\pm0.01, z=1.69\pm0.02, \theta=0.80\pm0.02$ and $\nu=1.45\pm0.05$.
%\deleted{It is clear that these exponents are similar to those of two special case of QKPZ (i) and (ii) mentioned above.} 
Furthermore, comparing our results to one of the discrete versions of QEW model \cite{song2008discrete}, we find consistency between them.
Secondly, we focus on QKPZ with $f=0$. For the (1+1)-dimensional case, numerical exploration was reported in \cite{ramasco2001interface}. Based on the scaling ansatzs mentioned in SEC. \ref{sec2}, we  achieve the critical state at $\lambda_c\approx2.865$ with system size $L\times L=256\times256$, and obtain the critical exponents $\delta=0.58\pm0.02, \beta=0.41\pm0.01, \theta=0.83\pm0.02, \nu=1.40\pm0.02, \alpha=0.51\pm0.01$ and $z=1.35\pm0.05$ in the (2+1)-dimensions. Comparing these scaling exponents with those calculated above, it can be seen that the universality class of this system is within the crossover from QKPZ to QEW. This is a trivial phenomenon, because when $f \to 0$, $\lambda$ will decrease. It is worth noting that the scaling exponent results at $f = 0$ are consistent with the directed percolation depinning (DPD) model \cite{Amaral1995Avalanches,family1992surface,barabasi1995fractal}. Ramasco et al. \cite{ramasco2001interface} proposed this conclusion for the (1+1)-dimensional case, and now we prove that it also holds in the (2+1)-dimensional case.

\begin{figure*}[htbp]
 \centering
        \begin{minipage}[t]{0.32\textwidth}
            \centering          %子图居中
            \includegraphics[width=54mm]{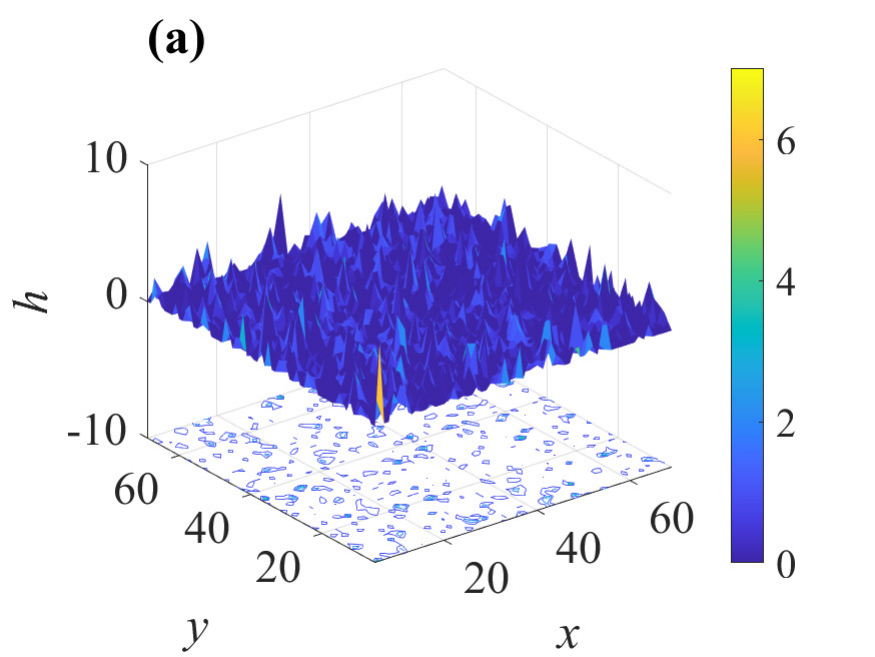} 
        \end{minipage}
        \begin{minipage}[t]{0.32\textwidth}
            \centering          %子图居中
            \includegraphics[width=54mm]{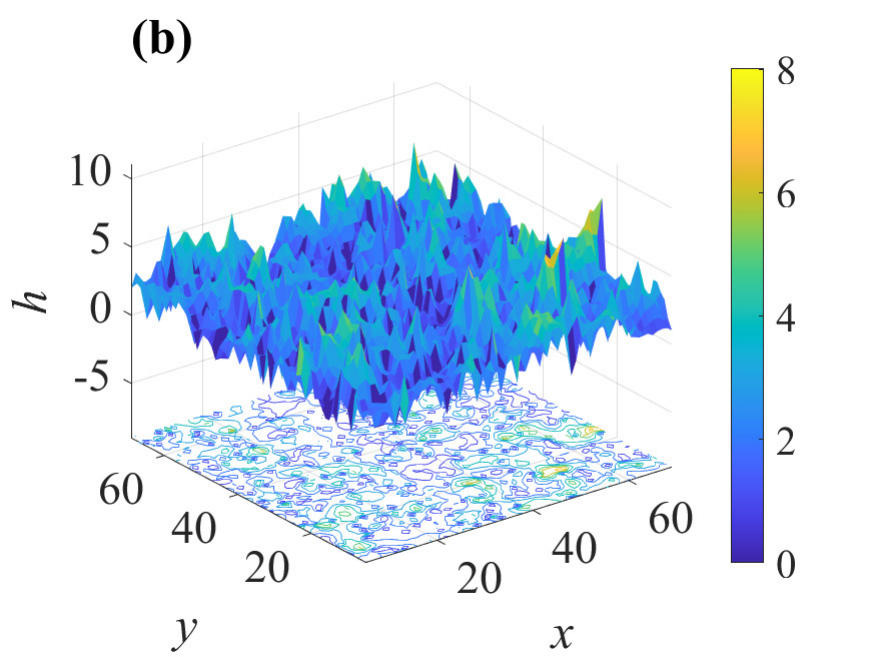} 
        \end{minipage}
        \begin{minipage}[t]{0.32\textwidth}
            \centering          %子图居中
            \includegraphics[width=54mm]{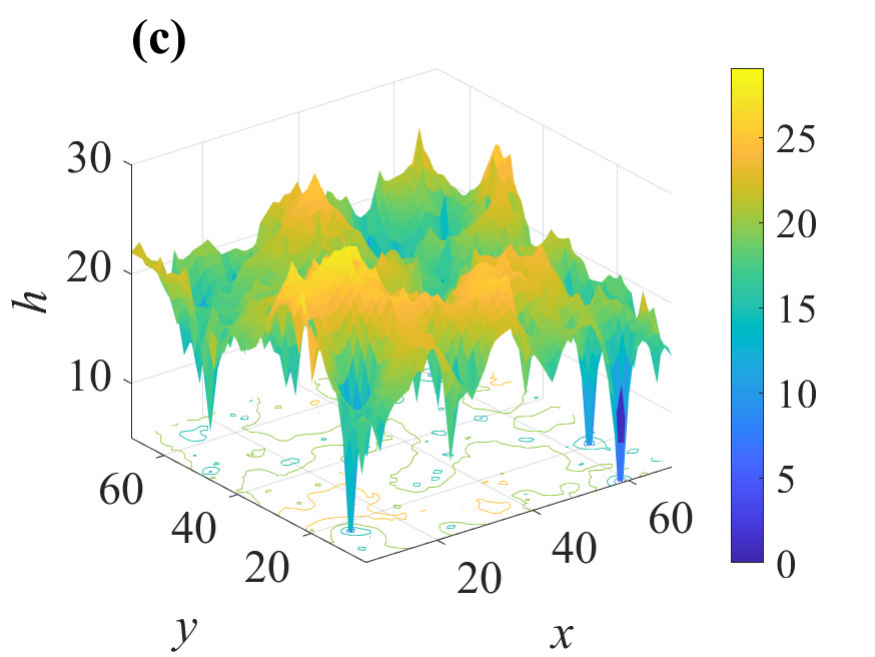} 
        \end{minipage}
  \\
         \begin{minipage}[t]{0.32\textwidth}
            \centering          %子图居中
            \includegraphics[width=54mm]{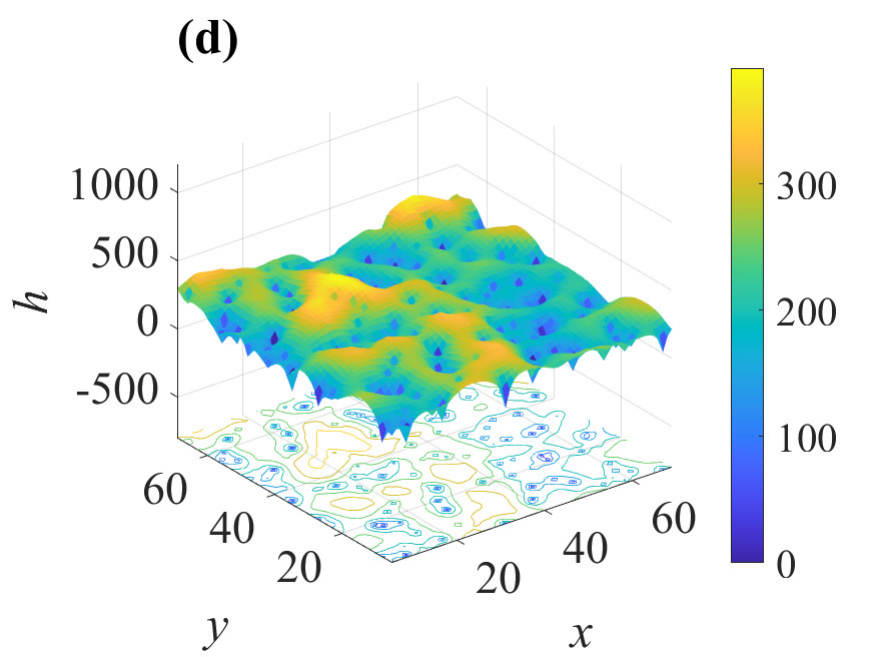} 
        \end{minipage}
        \begin{minipage}[t]{0.32\textwidth}
            \centering          %子图居中
            \includegraphics[width=54mm]{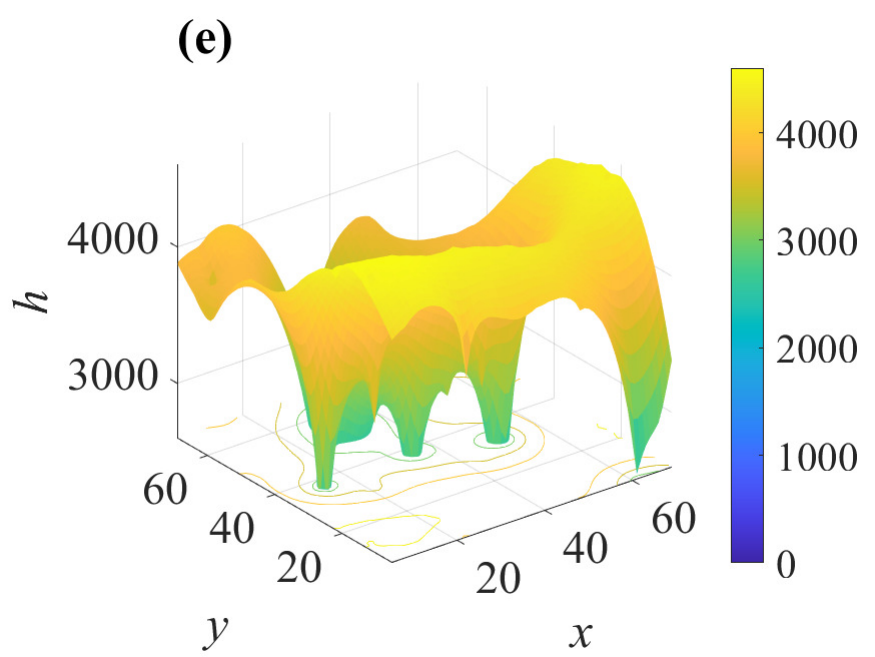} 
        \end{minipage}
        \begin{minipage}[t]{0.32\textwidth}
            \centering          %子图居中
            \includegraphics[width=54mm]{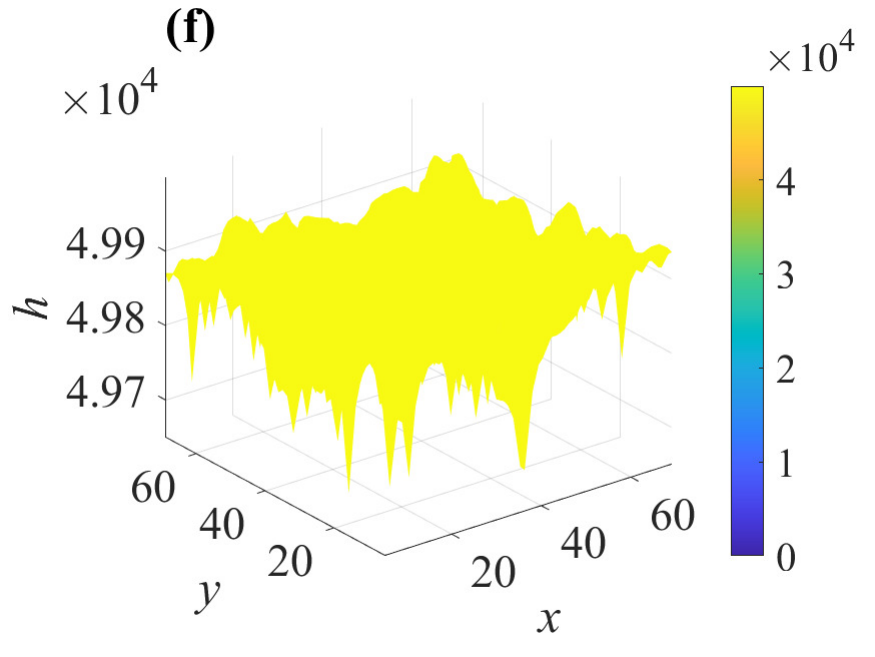} 
        \end{minipage}
   \\ 
   
        \begin{minipage}[t]{0.32\textwidth}
            \centering          %子图居中
            \includegraphics[width=54mm]{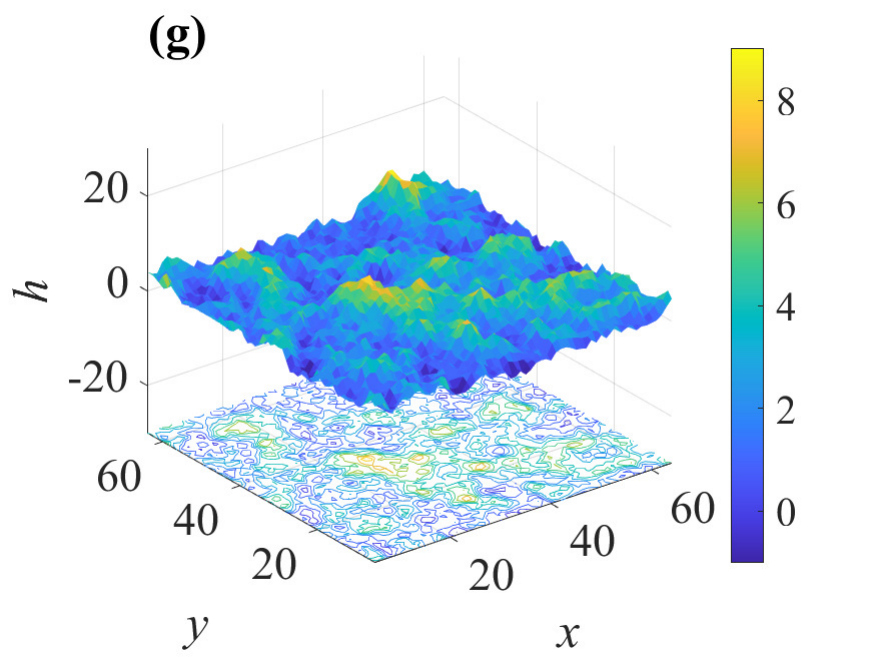} 
        \end{minipage}
        \begin{minipage}[t]{0.32\textwidth}
            \centering          %子图居中
            \includegraphics[width=54mm]{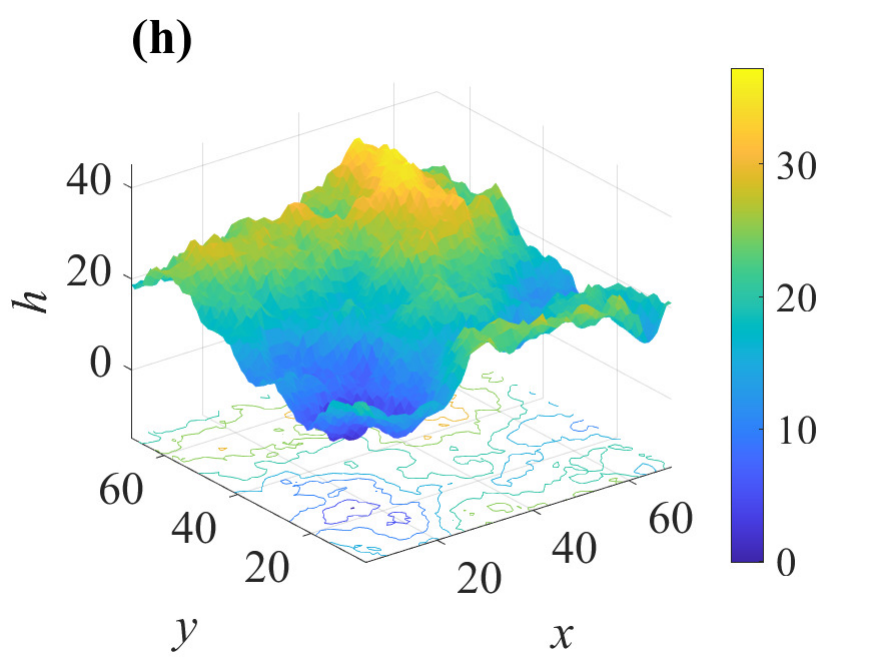} 
        \end{minipage}
        \begin{minipage}[t]{0.32\textwidth}
            \centering          %子图居中
            \includegraphics[width=54mm]{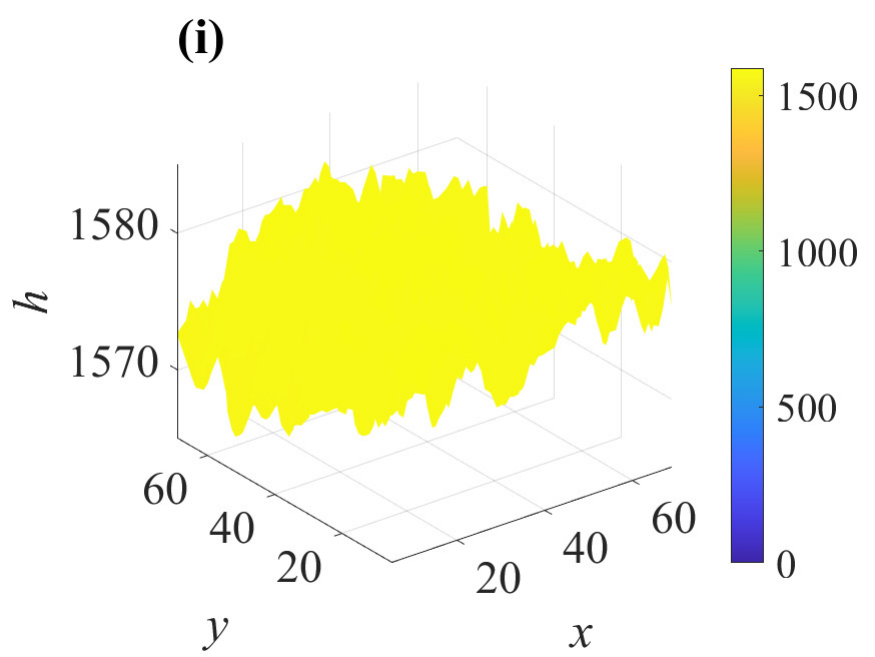} 
        \end{minipage}
    \\
        \begin{minipage}[t]{0.32\textwidth}
            \centering          %子图居中
            \includegraphics[width=54mm]{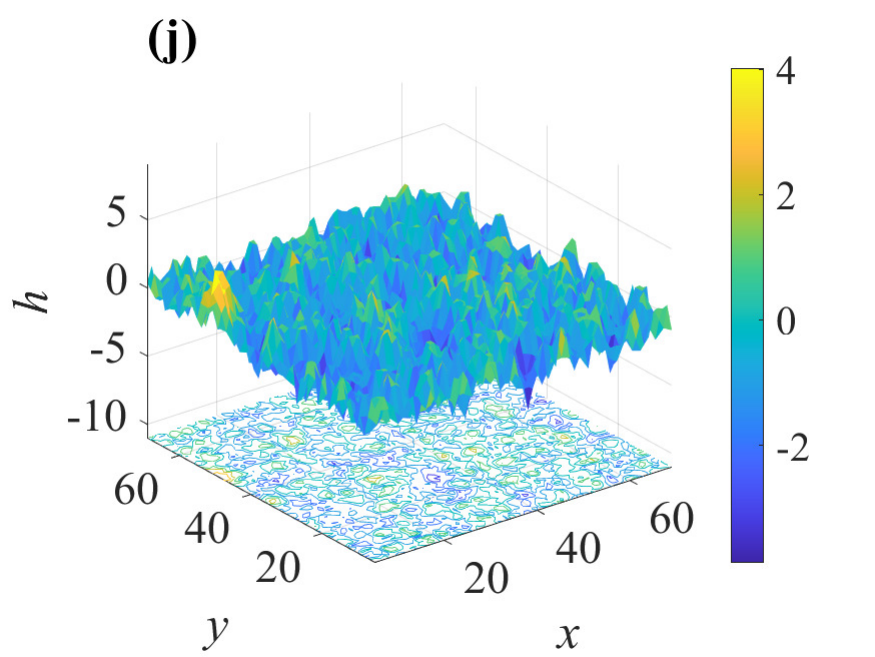} 
        \end{minipage}
        \begin{minipage}[t]{0.32\textwidth}
            \centering          %子图居中
            \includegraphics[width=54mm]{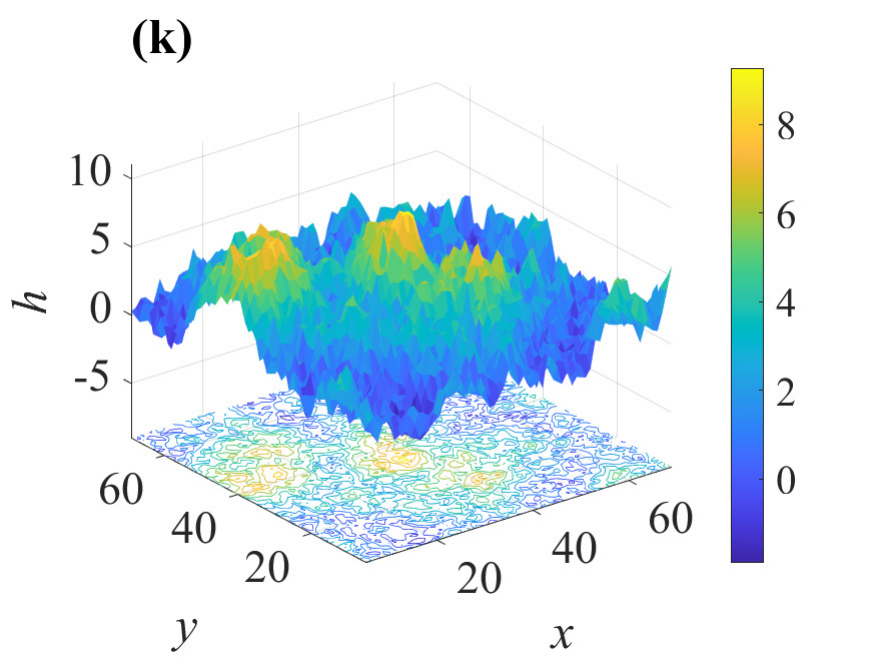} 
        \end{minipage}
        \begin{minipage}[t]{0.32\textwidth}
            \centering          %子图居中
            \includegraphics[width=54mm]{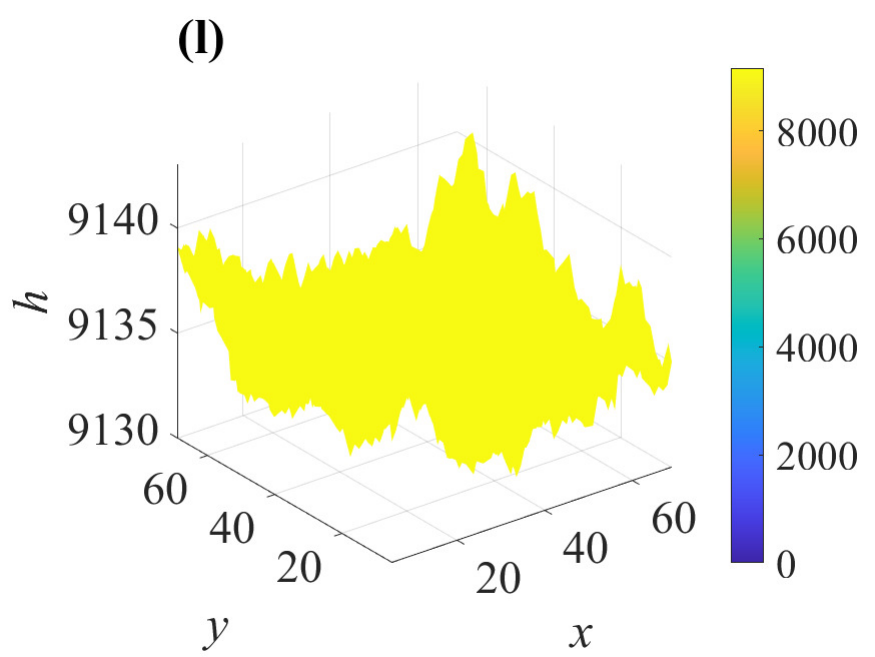} 
        \end{minipage}
    
     \caption{Surface morphologies in the (2+1)-dimensional { systems with quenched disorder} at $t=2^{18}$ with $\Delta t=0.01$. Color bar represents the local height of the interface. (a)-(c) in Model I, (d)-(f) in Model II, (g)-(i) in QEW, and (j)-(l) in QKPZ during the saturated time regimes with three different external driving forces including low $f$, $f_c$, and high $f$.}
     \label{fig13}
\end{figure*}
 
In order to visually give comparisons of {these systems with quenched disorder} in the (2+1)-dimensions, surface morphologies are presented with various external driving forces during the saturated time regimes, as shown in Figs. \ref{fig13}(a)-(l).  
When $f< f_c$, we show typical snapshots of the saturated surfaces at pinned region in the left column of Fig. \ref{fig13}, including four subgraphs: (a), (d), (g), and (j). In these cases, the velocities are almost zero since the surfaces are pinned, and we confirm these features by measuring average height field at different time instants. 
Typical snapshot configurations are obtained when $f=f_c$, as shown in the middle column of Fig. \ref{fig13}, including subgraphs: (b), (e), (h), and (k). It is evident that these surfaces undergo transitions from pinning to depinning phases near the critical thresholds for different {elastic-string} models. And surface morphologies change significantly with time evolution in comparison with those at pinning phases.
When  $f>f_c$, the evolving surfaces move forward prominently, and the average velocities remain almost unchanged for a certain model {with quenched disorder}. However, the magnitudes of velocities exhibit significant differences among these {elastic-string} models, primarily depending on the magnitudes of various parameters, including external forces and quenched noises.

\section{Discussions and conclusions}
\begin{table*}[ht]\footnotesize
\caption{\label{tab}Summary of scaling exponents at the critical pinning-depinning transitions in the (1+1)- and (2+1)-dimensions.}
%\begin{ruledtabular}
\begin{tabular}{clllllll}
\hline
Models     & Dimensions      & $\beta$  & $\delta$ & $\theta$  & $\nu $&$\alpha$& $z$\\
\hline
\multirow{3}{*}{Model I}& 1+1 \cite{biswas2011depinning}&$0.35\pm0.05$&$0.60\pm0.01$&$0.83\pm0.01$&$1.35\pm0.05$&-&-
\\
& 1+1 &$0.33\pm0.03$&$0.730\pm0.005$&$0.99\pm0.02$&$1.50\pm0.05$&$0.80\pm0.05$&$2.1\pm0.1$\\
& 2+1 &$0.120\pm0.005$&$0.900\pm0.005$&$1.03\pm0.05$&$1.48\pm0.03$&$0.55\pm0.03$&$4.0\pm0.3$\\
\hline
\multirow{3}{*}{Model II} &1+1 \cite{biswas2011depinning}&$0.65\pm0.05$&$0.34\pm0.01$&$1.00\pm0.01$&$2.95\pm0.05$&-&-
\\
&1+1&$0.585\pm0.003$&$0.375\pm0.003$&$0.368\pm0.003$&$0.98\pm0.03$&$0.95\pm0.05$&$1.5\pm0.1$\\
&2+1&$0.472\pm0.002$&$0.424\pm0.004$&$0.386\pm0.003$&$0.93\pm0.02$&$0.70\pm0.02$&$1.45\pm0.02$\\
\hline
\multirow{3}{*}{QEW}   &1+1 \cite{amaral1995scaling,kim2006depinning}&$0.85\pm0.03$& $0.128$&$0.24\pm0.03$&$1.96(6)$&$1.250(3)$&$1.440(15)$\\
 & 2+1 \cite{song2008discrete} &$0.493(3)$&$0.504(2)$&$0.631(2)$&$1.252(12)$&$0.755(5)$&$1.53(2)$\\
 & 2+1&$0.45\pm0.01$&$0.51\pm0.02$&$0.80\pm0.02$&$1.45\pm0.05$&$0.74\pm0.01$&$1.69\pm0.02$ \\
 \hline
\multirow{4}{*}{QKPZ} &  1+1 \cite{amaral1995scaling,lee2005depinning} &$0.67\pm0.05$&$0.360(1)$&$0.64\pm0.12$&$1.711$&$0.633(8)$&$0.978$\\
& 2+1 \cite{Mukerjee2022DepinningII}
&-&-&-&-&$0.4941$&$1.4112$\\
& 2+1 ($f=0$)&$0.41\pm0.01$&$0.58\pm0.02$&$0.83\pm0.02$&$1.40\pm0.02$&$0.51\pm0.01$&$1.35\pm0.05$\\
& 2+1 ($f\neq 0$)&$0.38\pm0.05$&$0.64\pm0.02$&$1.03\pm0.02$&$1.25\pm0.03$&$0.50\pm0.03$&$1.32\pm0.03$\\
\hline
\multirow{1}{*}{DPD} &  2+1 \cite{Amaral1995Avalanches} &$0.41\pm0.05$&-&-&-&$0.48\pm0.03$&-\\
\hline
\end{tabular}
%\end{ruledtabular}
\end{table*}

In this paper, we revisit two classes of discrete elastic-string models and generalize them to the (2+1)-dimensional cases to investigate the pinning-depinning transition. As the existing universality classes, the (2+1)-dimensional QKPZ and QEW equations are also studied numerically. Two discrete models {with quenched disorder} clearly exhibit the transition, especially, Model II shows high sensitivity to external parameters even in smaller system size. It should be emphasized that the results of Model I in the (1+1)-dimensional case differ from those of \cite{biswas2011depinning} since we set a different noise, which is equivalent to setting $p$ in \cite{biswas2011depinning} to 0. In Model II, we introduce term $\eta\left(\vec{r},h(\vec{r},t)\right)$ to replace 
term $\eta(\vec{r})$ in \cite{biswas2011depinning}. We argue that Model II driven by $\eta\left(\vec{r},h(\vec{r},t)\right)$ is more akin to the models describing crack front propagation \cite{demery2014microstructural,basu2019hydrodynamic,Maaloy1992Experimental,Elisabeth1997Scaling}. The comparisons between our estimated values and previous results are presented in Table \ref{tab}.

We adopt similar research methodologies across these { systems with quenched disorder}. Firstly, by observing a power-law decay of the interface velocity, we determine the critical threshold. Subsequently, the scaling exponents $\delta, \beta, \alpha, z, \theta$ and $\nu$ near the critical point are estimated based on various power-law scaling ansatzs. Our results show that the estimated values of these scaling exponents satisfy $\beta+\delta=1, z=\alpha/\beta$ and $\nu=\theta/\delta$ very well. All exponents calculated are tabulated in Table \ref{tab}, along with corresponding previous results. It should be noted that these discrete {elastic-string} models do not fall into the existing universality classes, including QEW and QKPZ systems.

For the QKPZ system, we generalize (1+1)-dimensional case to (2+1)-dimensions. Our simulations demonstrate that the pinning-depinning transition within the genuine QKPZ region can be achieved by adjusting parameters $f$ and $\lambda$, allowing for the determination of the scaling exponents.
Additionally, we calculate the critical exponents in the (2+1)-dimensional QEW, one of special case of QKPZ with $\lambda =0$, and the critical values obtained here are in line with those of the discrete model belonging to QEW universality class (the cellular automaton version of QEW) \cite{song2008discrete}.
Furthermore, we also focus on another special case of QKPZ system: $\lambda_c$ region with $f=0$. More specifically, the pinning-depinning transition in this special system can be determined by adjusting $\lambda$ when $f=0$. Near the critical threshold, the corresponding scaling exponents are between those of QEW and QKPZ universality classes to some extent,  and belong to the DPD class \cite{Amaral1995Avalanches,family1992surface,barabasi1995fractal}.
%\deleted{Through numerical simulations, we obtain similar scaling exponents by adjusting $F$ and $\lambda$ near their respective critical point based on Eqs. (\ref{eq15}-\ref{eq18}). It seems to indicate that the $\lambda$ terms in Eq. (\ref{eq10}) is equivalent to a constant that acts as an external driving force $F$ to some extent.} 
The detailed results and comparisons are presented in Table \ref{tab}.

The relationships between the correlation length of the models and the surface morphologies, as well as scaling exponents, are evident. According to the different forms of the elastic forces in Eqs. (\ref{eq5}),(\ref{eq6}),(\ref{eq8}), we know that Model I is a long-range model and Model II is a local one, which is beyond doubt. A similar comparison can be drawn between QKPZ and QEW. Referring to Fig. \ref{fig11}, the correlation function curve reaches saturation at a certain distance $r_c(t)$, where $r_c(t)$ is the correlation length \cite{jeong2002restricted} of the system at time $t$. We find $r^{QKPZ}_c(t)>r^{QEW}_c(t)$ at the same time, indicating that QKPZ possesses a longer correlation length compared to QEW. Now let us review Fig. \ref{fig13} with an eye toward the correlation length. The surface  morphologies of the  models with longer correlation length, i.e. Model I and QKPZ, are more finely fragmented, and the overall fluctuation of height is smaller. On the contrary, from the morphologies of the models with shorter correlation length, i.e. Model II and QEW, we find the overall fluctuation of height is larger. This disparity stems primarily from the fact that short-range correlation, in contrast to long-range correlation, fails to provide adequate synergistic effects at the interface. In addition, the influence of the nonlinear term in QKPZ on the surface morphology is evident: the lateral growth effect induced by the nonlinear term causes the larger peaks in the QEW morphology to fragment into smaller peaks in the QKPZ morphology. Besides, the change in the scaling exponents in Table \ref{tab} with the range of correlations shows the same trend in both pairs of models, implying some physical characteristics to be explored. We take the pair of Model I and Model II as an example to briefly illustrate these characteristics below. We find that, regardless of the dimension of the models, the growth exponent $\beta$ and roughness exponent $\alpha$ of the  model with longer correlation length (Model I) are always smaller than those of the model with shorter correlation length (Model II), and velocity exponent $\theta$ of the former is always greater than that of the latter. Based on scaling ansatz (\ref{eq14}), smaller $\beta$ and $\alpha$ mean that the interface roughness $W$ of the system with longer-range correlation grows more slowly in the early stage of evolution and the long-term saturation roughness $W(L,t\to \infty)$ is smaller. These differences can be attributed to stronger synergies between interface elements resulting from longer-range correlation. Considering Eq. (\ref{eq17}), we know that a larger $\theta$ means that the interface saturation velocity of a long-range system is more sensitive to changes in external forces. These rules are also applicable to the comparison between QKPZ and QEW. In addition, in (2+1)-dimensional case, we observe that the scaling exponents $\theta$ and $\alpha$ of Model I are close to those of QKPZ, suggesting analogous steady-state properties between the two models in the (2+1)-dimensional case.

The potential applications of these  {stochastic systems driven by quenched disorder} studied in current work deserve further investigations. The relationship between these elastic-string models and fracture front propagation is an interesting topic \cite{demery2014microstructural,bonamy2011failure,ponson2009depinning,le2020universal,maaloy2006local,tanguy1998individual,ponson2010crack,alava2006statistical,gao1989first}. The different properties of elastic force of these two quenched models correspond to different materials, such as ductile and brittle materials \cite{demery2014microstructural,bonamy2011failure,ponson2010crack,alava2006statistical}. Specifically, the roughness exponent of (1+1)-dimensional Model I is consistent with that of the material Al-Si alloy AA4253 ($\alpha=0.82$) \cite{Maaloy1992Experimental} and the ductile fracture ($\alpha=0.8$) \cite{Bouchaud1990Fractal}. The roughness exponent of (1+1)-dimensional Model II is consistent with that of the brittle material plaster of Paris ($\alpha=0.95$) \cite{Maaloy1992Experimental}. Furthermore, the results of our Model I and Model II align with the assumption of universality ($\alpha=0.87\pm 0.07$) \cite{Maaloy1992Experimental,Elisabeth1997Scaling}. Brittle materials may exhibit a larger roughness exponent \cite{Maaloy1992Experimental}. It is more appropriate to state that the smaller the correlation length is, the larger the roughness exponent is \cite{Elisabeth1997Scaling}. This rule is in good agreement with these two models mentioned here. Elastic description is also related to self-organized critical phase transition in some special cases \cite{bonamy2008crackling,laurson2016Interevent}. And elastic energy \cite{ponson2016statistical} in elastic-string model for fracture is worth studying.  Besides, as mentioned in SEC. \ref{sec1}, the wetting fronts \cite{le2009height,barabasi1995fractal,moulinet2004width,Roux2003effective,Joanny1984a,Buldyrev1992Anomalous} in disordered materials are relevant to the (2+1)-dimensional QKPZ with $f=0$. For instance, in a wetting cubic sponge without external pressure, the interface could propagate in quenched random media driven by local capillary forces, implying that pinning-depinning transitions occur even in the absence of external driving forces. Furthermore, exploring the confining potential-driven interface \cite{Wiese2022,mukerjee2022depinning,Mukerjee2022DepinningII} is a promising research direction. This exploration would aid in constraining and constructing field theories, facilitating a deeper understanding of the statics and dynamics of the {elastic-string} systems  that we are interested in. 
In addition, it is important to note that, although these two discrete elastic-string models exhibit a rich variety of dynamic scaling properties near the critical threshold, and the estimated values of critical exponents are in agreement with some specific experimental results, the physical meanings for the elastic interaction term are still not very clear to explain effectively the concrete physical properties of different materials. { Further investigations into a more profound connection between the elastic-string models and the related experimental results would be of interest and potential value.}
%For these elastic-string models studied in current work, the noise interval is asymmetric. When one considers a symmetrical noise interval, it is possible that there may obtain some interesting results corresponding to \cite{mukerjee2022depinning,Mukerjee2022DepinningII}. Further investigation of this issue would be of interest and value.}

\section*{Acknowledgments}

We would like to thank Juemin Yi and Yueheng Lan for useful discussions and critical reading of the manuscript. This work is supported by Key Academic Discipline Project of China University of Mining and Technology (CUMT) under Grant No. 2022WLXK04, and Undergraduate Training Program for Innovation and Entrepreneurship of CUMT under Grant No. 202110290059Z.

%%end novalidate
\section*{References}
\bibliography{ref}

\end{document}